\documentclass[aip,apl,reprint,amsmath,amssymb]{revtex4-2}
\usepackage{graphicx}
\usepackage{booktabs}
\usepackage{xcolor}
\usepackage[hidelinks]{hyperref}

\newcommand{\um}{\,\mu\mathrm{m}}

\begin{document}

\title{An Agent-Driven, Cross-Validated Inverse-Design Pipeline for Dielectric Metasurfaces}

\author{Giuseppe Emanuele Lio}
\email{giuseppe.lio@unipi.it}
\affiliation{Department of Physics ``E. Fermi'', University of Pisa, Largo B. Pontecorvo 3, 56127 Pisa, Italy}

\author{Roberto Caputo}
\affiliation{Department of Physics, University of Calabria, 87036 Rende (CS), Italy}
\affiliation{Consiglio Nazionale delle Ricerche - Istituto di Nanotecnologia (CNR-Nanotec), Rende (CS), 87036 Italy}

\author{Alain J. Corso}
\affiliation{Consiglio Nazionale delle Ricerche, Istituto di Fotonica e Nanotecnologie (CNR - IFN), Via Trasea 7, 35131 Padova, Italy}

\author{Maria Guglielmina Pelizzo}
\email{pelizzo@dei.unipd.it}
\affiliation{Dipartimento di Ingegneria dell'Informazione, Università degli Studi di Padova, Via G. Gradenigo 6/B, 35131 Padova, Italy}
\affiliation{Consiglio Nazionale delle Ricerche, Istituto di Fotonica e Nanotecnologie (CNR - IFN), Via Trasea 7, 35131 Padova, Italy}

\author{Antonio Ferraro}
\affiliation{Consiglio Nazionale delle Ricerche - Istituto di Nanotecnologia (CNR-Nanotec), Rende (CS), 87036 Italy}

\begin{abstract}
Inverse design of dielectric metasurfaces is powerful but labour-intensive: each new target requires choosing a parameterization and a figure of merit, managing compute resources, and critically assessing results that may be numerically unconverged, sensitive to fabrication errors or trapped in local optima. Here we present an open-source pipeline in which an artificial intelligence (AI) coding agent turns a plain-language specification into metasurface optimization runs, executes them, analyses every run with a fixed checklist and proposes the next iteration. Designs are optimized with differentiable three-dimensional rigorous coupled-wave analysis (RCWA), including robustness to fabrication tolerances, and every reported design is independently verified with finite-difference time-domain (FDTD) simulations. We demonstrate the workflow at 1550 nm on a freeform silicon metagrating, on a polarization-independent meta-atom library
assembled into beam-deflecting supercells, and on metalenses built from the same library. In the metagrating case, the independent FDTD
verification revealed a 5\% overestimate of the efficiency by the optimization model that a conventional convergence test had not detected, and the robust design retained most of its efficiency under $\pm$10~nm fabrication errors. In the other cases, instead, the obtained RCWA results are confirmed by full-wave simulations.
\end{abstract}

\maketitle

Dielectric metasurfaces can shape the phase, amplitude and polarization of light with sub-wavelength scatterers \cite{Yu2014, Arbabi2015}. They are designed either by assembling libraries of characterized meta-atoms, as in blazed binary gratings and metalenses \cite{Lalanne1998, Khorasaninejad2016, Lalanne2017, padovani2025blazed}, or by inverse design of the whole device\cite{Molesky2018, Jensen2011, yesilyurt2023fabrication, chen2026llm}. Differentiable implementations of rigorous coupled-wave analysis (RCWA) \cite{Moharam1981, Moharam1995} built on automatic-differentiation frameworks \cite{Colburn2021, Kim2023,Paszke2019} have made such optimizations fast, and produce metagratings and metalenses with efficiencies well beyond phase-sampled designs \cite{Sell2017, Jiang2019, Pestourie2018, Chung2020}. The difficulty has therefore shifted from computing an optimal design to deciding whether the result can be trusted. An optimized design may lose much of its efficiency under nanometre-scale fabrication errors~\cite{Wang2011,Hammond2021}, may be only a local optimum, or may owe part of its computed efficiency to an unconverged model, for instance too few Fourier harmonics in RCWA. These problems are well known, but checking them for every run is tedious and often skipped.

AI agents that write code, run tools and interpret results have begun to automate the research loop in chemistry \cite{Boiko2023, Bran2024}. In photonic design, AI has so far been used in three
roles: (i) as a scripting assistant, (ii) as a fast replacement of the electromagnetic solver, and (iii) as an agent that drives a rigorous solver. In the first role, language models \cite{Kim2025LLM, Lu2025} write and explain simulation code but leave verification to the user. In the second, generative models and neural surrogates \cite{Ma2024, Jiang2021} replace the
solver; in MetaChat \cite{Lupoiu2025}, for example, a multi-agent system drives a surrogate of Maxwell's equations to design two-dimensional, single-polarization devices in minutes, with full-wave checks of selected designs. This speed comes at the cost of an accuracy bounded by the training data. The present work belongs to the third role: the agent runs a rigorous three-dimensional solver in the loop and, bound by a written protocol, every reported result is verified by an independent method.

In the following, We describe the workflow and apply it to three case studies operating at a wavelength of 1550~nm: a freeform silicon metagrating, a polarization-independent meta-atom library assembled into supercells that deflect light at a chosen angle, and a metalenses built from the same library. 

\begin{figure}[t]
\includegraphics[width=\linewidth]{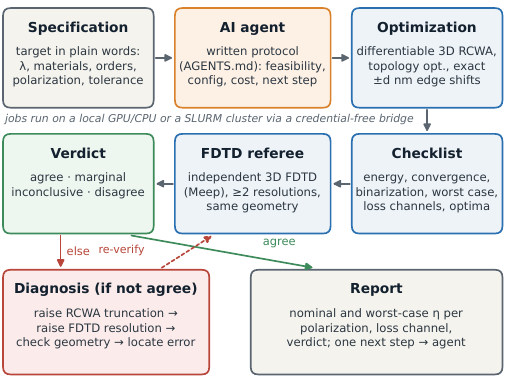}
\caption{Agentic design loop with an independent verification. Every reported design passes a fixed checklist and a 3D
FDTD re-simulation; a verdict other than \emph{agree} triggers a prescribed diagnosis before anything is reported.}
\label{fig:loop}
\end{figure}

Figure~\ref{fig:loop} summarizes the loop. The user defines a target in plain language, e.g., ``maximize the +1 transmitted order at 1550~nm for unpolarized light, silicon on silicon, robust to $\pm$10~nm fabrication error''. A general-purpose coding agent follows a written protocol, stored in the repository as
\texttt{AGENTS.md}; because the protocol is written down, the agent's behavior is documented and reproducible. The general-purpose agent first checks physical feasibility (e.g., which diffraction orders propagate and whether the thickness gives enough phase range), then writes the configuration file, estimates the computational cost in GPU or CPU hours and asks for approval before computationally expensive runs. Jobs are submitted through a minimal file-based bridge. The agent never connects to the computing resources itself: it only writes one-line requests to a queue folder. A watcher process, started
by the user in their own terminal after logging in to the cluster, checks each request against a short list of allowed commands and executes it on the local GPU or on the HPC cluster. The user therefore performs all authentication, including two-factor login, and the agent never handles passwords or keys (Sec.~S1 of the Supplementary Material).

The design itself is produced by a standard gradient-based topology optimization: the agent prepares and launches each run using RCWA solver, but the update of the design at every iteration is computed by the optimization algorithm, not by the agent. A single patterned layer is described by a material density $\rho \in [0,1]$ (1 = silicon, 0 = air) on a square pixel grid; every pixel is a design variable, so the layout is \emph{freeform}, not restricted to predefined shapes such as pillars, holes or bars. The pixel size is chosen for each problem; in all the three case study hereafter reported it is 5~nm, small compared with the $\pm10$~nm fabrication tolerance and with the filter radii of the smoothing filters (25 and 40~nm). This geometric grid is independent of the number of Fourier harmonics used by RCWA, which sets the accuracy of the electromagnetic calculation. TORCWA~\cite{Kim2023}, written in PyTorch~\cite{Paszke2019}, computes on the GPU the scattering matrix of the layer, and automatic differentiation gives the derivative of the figure of merit (FOM) with respect to all pixels in one additional backward pass. The optimizer Adam~\cite{Kingma2015}, developed for training neural networks, uses these derivatives to update the density at every iteration. Two further operations make the design manufacturable. A Gaussian filter (radius $R = 40$~nm) smooths the
density and removes features much smaller than this radius. During the optimization, however, pixels can take intermediate values, which do not correspond to a real material. The filtered
density is therefore passed through a smooth step function, a hyperbolic tangent centered at 0.5, which pushes each pixel towards 0 (air) or 1 (silicon) while remaining differentiable. Its steepness $\beta$ starts low, so that the optimizer can still reshape the design, and is increased gradually until the pattern is binary \cite{Wang2011}. 
For the metagrating, the FOM is the +1-order efficiency for unpolarized light, $\mathrm{FOM} = \frac12\sum_{i\in\{x,y\}} \eta^{(i)}_{+1}$, where $\eta^{(i)}_{+1}$ is the fraction
of the incident power transmitted into the +1 order for input light polarized along $i$, including both polarization components (s and p) of the diffracted light.
Fabrication robustness is imposed geometrically. At every iteration the projected, nearly binary design is blurred with a second Gaussian kernel, $\exp(-r^2/R_2^2)$ with $R_2 = 25$~nm, which turns a straight edge into an error-function profile; thresholding this profile at $t_{e,d} = \tfrac12[1 \pm \mathrm{erf}(d/R_2)]$ therefore moves every straight edge inwards (eroded design) or outwards (dilated design) by exactly $d = 10$~nm. We call this an edge shift of $\pm d$; it mimics fabricated features that are narrower or wider than designed. The eroded, nominal and dilated designs are simulated together and combined with a smooth minimum, $\mathrm{FOM}_\mathrm{rob} = -k^{-1}\ln\sum_j e^{-k\,\mathrm{FOM}_j}$ with $k = 10$, where $\mathrm{FOM}_j$ is the FOM of design $j$; $\mathrm{FOM}_\mathrm{rob}$ is the FOM of the robustness which approximates the worst case and is the quantity maximized \cite{lio2025design}. Unlike the classical threshold formulation \cite{Wang2011}, in which the shift depends on the local grey level, this enforces exactly the requested tolerance. Every binary design is then re-evaluated with exact edge shifts from Euclidean distance transforms, and ranked by its worst case (see sec. S2 of Supplementary Material).

Each run is analyzed with the same checklist: energy balance, convergence at a higher Fourier truncation, binarization, worst case, minimum features, loss channels per polarization and local optima (see sec. S3 of Supplementary Material). Designs that pass the checklist are re-simulated in three dimensions with Meep FDTD solver \cite{Oskooi2010} at two or more resolutions, with the efficiency of each diffraction order obtained by projection onto the s- and p-polarized diffracted plane waves. A report assigns a verdict per design and polarization: \emph{agree}, if the target efficiency and the total transmission differ by less than 3\% and 2\%; \emph{inconclusive}, if the FDTD value is not yet resolution-converged; \emph{marginal}, within twice the tolerance; \emph{disagree} otherwise. For any verdict other than \emph{agree} the protocol prescribes a diagnosis ladder (raise the RCWA truncation, raise the FDTD resolution, verify that both solvers see the same geometry) before a number may be reported (see sec. S4 of Supplementary Material).

\begin{figure*}[ht]
\includegraphics[width=0.78\linewidth]{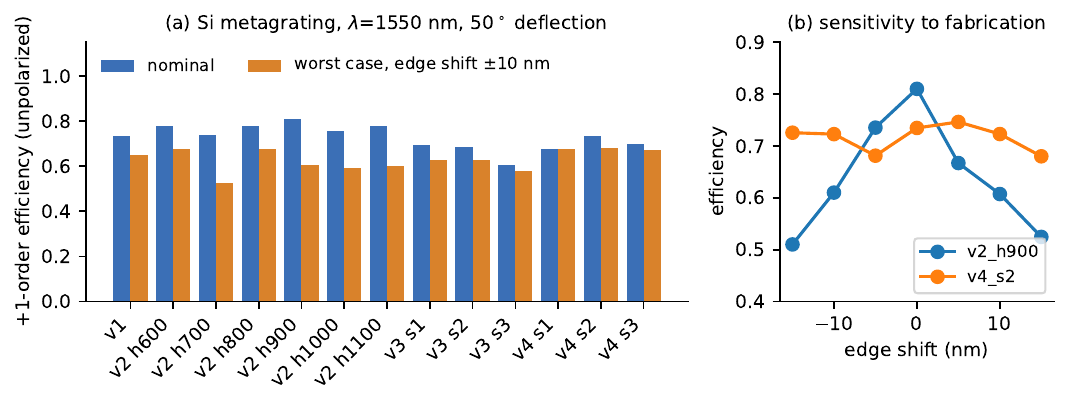}\\[2pt]
\begin{minipage}{0.78\linewidth}\small (c)\\ \includegraphics[width=\linewidth]{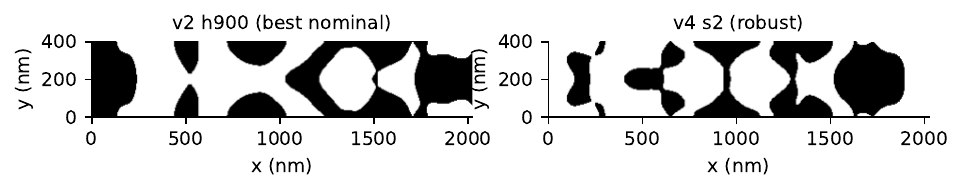}\end{minipage}
\caption{Silicon metagrating deflecting unpolarized 1550 nm light by 50$^\circ$. (a) Nominal +1 efficiency (blue) and worst case with edge shifts of $\pm$10 nm  (orange) for all design rounds: first run (v1), thickness sweep (v2, $h$ in nm), threshold-based robust optimization (v3) and geometric robust optimization (v4, three seeds). (b) Efficiency versus edge shift for the best nominal (v2 h900) and the best robust design (v4 s2); positive shifts dilate the silicon. (c) Freeform unit cells of v2 h900 (left) and v4 s2 (right), $h = 900$~nm; black is silicon.}
\label{fig:grating}
\end{figure*}

The two methods (RCWA and FDTD) approximate Maxwell's equations in different ways and therefore make different errors. RCWA expands the permittivity of the patterned layer and the electromagnetic fields in Fourier series along the two periodic directions and keeps only a finite number of terms (harmonics). We denote this truncation by $[N_x,N_y]$: the field is described as a sum of plane waves, one for each pair of harmonic orders $(m,n)$ with $|m| \le N_x$ and $|n| \le N_y$, each traveling in a direction fixed by the periods of the cell. A truncation of $[25,5]$ therefore uses $(2\cdot25+1) \times (2\cdot5+1) = 51 \times 11 = 561$ plane waves. The result approaches the exact solution as $N_x$ and $N_y$ increase, while the computational cost grows roughly as the cube of the number of plane waves; convergence is slow for abrupt silicon--air boundaries~\cite{Li1996}. FDTD instead discretizes space on a grid of small cubic cells: inclined or curved boundaries are approximated by steps of one cell (stair-casing), and the open space above and below the structure is truncated by absorbing layers that reflect a small fraction of the outgoing light \cite{Taflove2005, Oskooi2010}. Since these errors have independent origins, agreement between the two methods is strong evidence that a result is correct.

The first user request was a transmissive freeform silicon metagrating ($n = 3.476$) on a silicon substrate, with air above, that deflects unpolarized, normally incident light into the +1 order at 50$^\circ$. The rectangular unit cell has periods $\Lambda_x = 2023$~nm and $\Lambda_y = 400$~nm: $\Lambda_x$ follows from the grating equation, $\sin\theta = \lambda/\Lambda_x$, for $\theta = 50^\circ$ in air, whereas $\Lambda_y$ is shorter than the wavelength in silicon, so that no diffraction order propagates along $y$ and the light is deflected only in the $xz$ plane. The optimization used a truncation of $[25,5]$; one simulation together with its gradient took about 1.3~s on one NVIDIA V100S GPU of the CINECA cluster. A first run reached $\eta_{+1} = 74\%$; a sweep of the silicon thickness $h$, proposed by the agent, gave the best nominal design, $h = 900$~nm with $\eta_{+1} = 81\%$ [Fig.~\ref{fig:grating}(a)]. This design, however, is among the most fragile: its efficiency falls to $61\%$ for edge shifts of $\pm10$~nm and to $51\%$ for $\pm15$~nm [Fig.~\ref{fig:grating}(b)]. A first robust optimization used the classical threshold formulation~\cite{Wang2011}, in which the eroded and dilated designs are obtained by centreing the step function at 0.6 and 0.4 instead of 0.5. With these values the edges moved by 17--20~nm instead of the expected 5~nm, so the optimizer protected the design against errors much larger than intended and sacrificed nominal efficiency. Instead, the geometric formulation, which enforces the requested $\pm10$~nm exactly, produced a design whose efficiency is almost flat up to $\pm15$~nm, with a worst case of $\eta_{+1} = 68\%$ within $\pm10$~nm, the highest of all runs, at the cost of 7.5\% of nominal efficiency. The robust layout avoids the narrow bridges and gaps of the nominal optimum [Fig.~\ref{fig:grating}4(c)]. The analysis of all diffraction orders shows that the efficiency is limited mainly by reflection back into the high-index substrate, with less than 2\% of the power in other transmitted orders. Reflection is about 18\% for x-polarized and 34\% for y-polarized input, the latter close to the 31\% reflected by a flat silicon--air interface at normal incidence; an anti-reflection layer or a two-layer design is the natural next step (see the full design history in sec. S5 of Supplementary Material).


The usual convergence check is to recompute the final design at a higher truncation and verify that the FOM changes little. For the robust design v4~s2, the design with the highest worst-case efficiency, raising the truncation from $[25,5]$ to $[35,7]$ changed the FOM from 74.3\% to 73.5\%, a relative change of 1.1\%, below the 2--3\% usually accepted in RCWA
studies.

\begin{figure}[t]
\includegraphics[width=\linewidth]{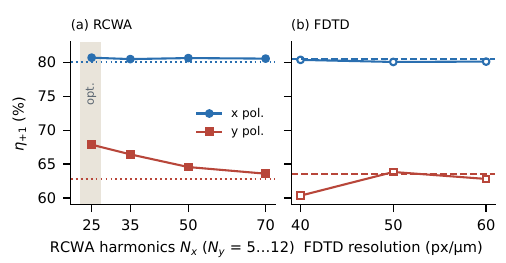}
\caption{Diagnosis of the RCWA - FDTD disagreement for the robust metagrating. (a) RCWA +1 efficiency versus the
number of Fourier harmonics $N_x$ ($[N_x,N_y]$ = $[25,5]$, $[35,7]$, $[50,10]$, $[70,12]$; shaded: truncation used in
the optimization); dotted lines: FDTD at 60 px/$\mu$m. (b) FDTD efficiency versus resolution; dashed lines: RCWA at
$[70,12]$. Only y polarization (red) is affected.}
\label{fig:trunc}
\end{figure}

The independent FDTD simulation, however, did not confirm the RCWA result for y polarization (Fig.~\ref{fig:trunc}). We quantify the agreement by the difference between the efficiencies computed by the two methods for the same design and input polarization, $\Delta\eta = \eta^\mathrm{RCWA}_{+1} - \eta^\mathrm{FDTD}_{+1}$; the protocol requires $|\Delta\eta| \le 3\%$ for agreement. For x-polarized input $|\Delta\eta| \le 0.7\%$ at all truncations and resolutions. For y-polarized input $\Delta\eta = 5.1\%$ at the optimization truncation $[25,5]$ and $3.6\%$ at $[35,7]$; the second value exceeds the 3\% tolerance, and the comparison was classified as \emph{marginal}. Following the diagnosis prescribed by the protocol, the agent increased the RCWA truncation. The y-polarized efficiency decreased steadily towards the FDTD value, whereas the x-polarized one did not change [Fig.~\ref{fig:trunc}(a)]; at $[70,12]$ harmonics $\Delta\eta = 0.8\%$, and the comparison was classified as \emph{agree} (Table~\ref{tab:valid}).

The error therefore lay in the computed efficiency, not in the design, and affected only one polarization. This is consistent with the slow convergence of the Fourier expansion for field components normal to high-index-contrast boundaries \cite{Li1996}. The freeform cell is patterned along both axes [Fig.~\ref{fig:grating}(c)]: the necks and lateral edges of the silicon features vary along $y$, where the optimization used only 5 harmonics, and the y-polarized electric field is normal to these silicon--air walls, so its discontinuity is poorly represented by the truncated expansion. Increasing $N_y$ together with $N_x$ removes most of the overestimate, and Fourier factorization rules~\cite{Li1996} would remove it at a lower cost.

This has two implications for automated optimization, where the model is trusted at every iteration. First, a small change between two neighboring truncations does not prove convergence when the convergence is slow: here each increase of the truncation lowered the y-polarized efficiency by only 1--2\%, but the accumulated overestimate at the optimization truncation was 5\%. Second, an optimizer maximizes the efficiency computed by its model, so it can favor geometries for which the numerical error happens to increase that efficiency; from the model alone it is impossible to tell whether this has occurred. Here the overestimate was moderate and the robust design was confirmed, but only the independent solver could show it: relying on its own model, the agent would have reported $\eta_{+1} = 67.9\%$ for y polarization instead of the 63--64\% given by convergence between RCWA and FDTD.

\begin{table}[b]
\caption{Cross-solver validation of the final designs. Gratings and supercell: $\eta_{+1}$ from RCWA at the highest truncation ($[70,12]$ and $[84,7]$) and from FDTD at the highest resolution (60 and 50 px/$\mu$m); all
verdicts \emph{agree}. Metalenses: focusing efficiency and depth of focus from the locally periodic RCWA (LPA) and from 3D FDTD (30 px/$\mu$m) propagated with the same code.}
\label{tab:valid}
\begin{ruledtabular}
\begin{tabular}{llcc}
Design & Opt. Prop. & RCWA & FDTD \\
\hline
Metagrating v4 s2   & $\eta_{+1}$, x pol. (\%)  & 80.6 & 80.1 \\
                    & $\eta_{+1}$, y pol. (\%)  & 63.6 & 62.8 \\
Supercell, \\ N. of meta-atoms 12  & $\eta_{+1}$, x pol. (\%)  & 88.1 & 89.2 \\
                    & $\eta_{+1}$, y pol. (\%)  & 92.5 & 90.9 \\
Metalens, NA 0.50   & focusing eff. (\%)        & 81.7 & 77.3 \\
                    & DOF ($\mu$m)              & 9.0  & 9.1  \\
Metalens, NA 0.25   & focusing eff. (\%)        & 88.4 & 90.1 \\
                    & DOF ($\mu$m)              & 25.7 & 25.2 \\
\end{tabular}
\end{ruledtabular}
\end{table}
The agent-driven protocol scales from single unit cells to devices assembled from a library of meta-atoms. 

In a second request, the user asked for polarization-independent beam-deflecting silicon meta-atoms on glass (SiO$_2$) in a square lattice of period $P_x=P_y = 700$~nm. A library-based device needs meta-atoms whose transmitted phase covers the full $0$--$360^\circ$ range. The agent's feasibility check showed that the originally proposed thickness $h$ of 100~nm covers only about 50$^\circ$ of phase; a sweep of transmission and phase of the silicon disks versus radius and height selected $h = 800$~nm, the smallest thickness that covers $360^\circ$ with high transmission.

Each phase level of the library is a separate optimization with its own target phase, and the square symmetry imposed on each meta-atom ($D_4$) makes its response independent of polarization. In a first attempt with  eight phase levels, all optimizations started from the same disk (radius 300~nm), only the target phase being different. Five of the eight ended in local optima, far from their target phase. Following the checklist item on local optima, the agent restarted each level from a clover and a cross of equal area, keeping the best result. With this initialization the final library was built with 24 phase levels, robust to $\pm$10~nm edge shifts; 20 of them transmit more than 95\% of the incident power (see sec.~S6 of the Supplementary Material).

A supercell of $N$ library atoms with phases increasing in steps of $360^\circ/N$ deflects light into the +1 order at $\sin\theta = \lambda/(NP)$. Supercells with $N = 8$, 12 and 24 reach $\eta_{+1} = 87$--$90\%$ for unpolarized light. A global robust optimization of the whole 12-atom supercell, started from the library assembly, raised its worst-case efficiency for edge shifts up to $\pm$10~nm from 75\% to 85\%, at the cost of 1.4\% of nominal efficiency [Fig.~\ref{fig:lens}(a) and sec.~S7 of the Supplementary Material]; for this design $|\Delta\eta| \le 1.6\%$ for both polarizations (Table~\ref{tab:valid}).

\begin{figure*}[t]
\includegraphics[width=\linewidth]{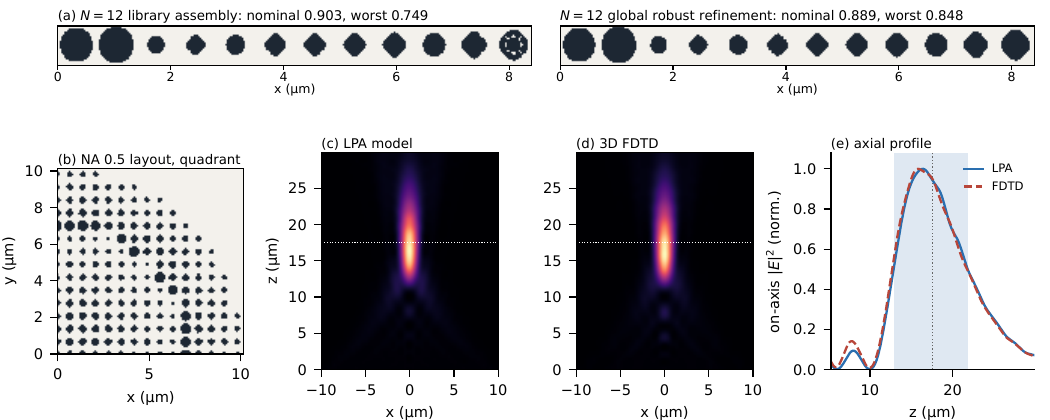}
\caption{Devices assembled from the meta-atom library ($h = 800$ nm Si on SiO$_2$, $P_x=P_y = 700$ nm; dark: silicon). (a) 12-atom supercell deflecting into the +1 order at 10.6$^\circ$, as assembled from the library (left) and after global robust refinement (right), with nominal and worst-case (edge shifts of $\pm$10 nm) efficiency for unpolarized light. (b) One quadrant of the NA 0.5 metalens ($D = 20.3\um$, $29\times29$ atoms); the optical axis is at the origin. (c,d) $|E|^2$ in the $xz$ plane from the RCWA and the FDTD near field, both propagated with
the same vector angular-spectrum code and normalized to their maxima; dotted line: design focal length. (e) On-axis intensity; shaded band: depth of focus of the model.}
\label{fig:lens}
\end{figure*}

The library can be also used to define aperiodic devices. Two circular metalenses of $D = 20.3\um$ were assembled by choosing, for each cell, the atom that maximizes $\sqrt{T_k}\cos(\phi_k - \phi)$ [Fig. \ref{fig:lens}(b)], and numerically designed with the locally periodic approximation \cite{Pestourie2018} and vector angular-spectrum propagation, which takes seconds (see sec. S8 of Supplementary Material). The whole apertures and near field propagation were then simulated with 3D FDTD, so that every difference comes from the near field alone [Fig. \ref{fig:lens}(c)--(e)]. At an NA of 0.5 the focus is diffraction limited (FWHM 1.60/1.43 $\mu$m in $x/y$ from FDTD, limit 1.59 $\mu$m), the depth of focus agrees within 0.1 $\mu$m, and the focusing efficiency is 77.3\% in FDTD against 81.7\% in the locally periodic RCWA; The 4\% gap is attributed to near-field coupling: the library characterizes each atom among identical neighbours, whereas in the lens its neighbours differ and their overlapping near fields
modify its transmission. The effect is strongest at the lens edge, where the phase changes fastest (about 4.4 atoms per $2\pi$), consistent with the supercell results. At an NA of 0.25, FDTD gives a slightly higher efficiency (90.1\% versus 88.4\%), and both methods place the intensity maximum well before the geometric focus, as expected for a Fresnel number of 1.7 \cite{LiWolf1981}. The 0.25 NA metalens is shown in sec. S8 of Supplementary Material.

To test whether the agent-driven protocol carries over to problems outside the previous case studies, a new agent session, started without prior context, was asked to design a TM quasi-bound state in the continuum (quasi-BIC) with $Q \approx 1000$ near 1550~nm in Si on SiO$_2$ (see sec.~S12 of Supplementary Material). Because the pipeline had no objective for the quality factor, the agent added a spectral scan with a Fano fit based on the same RCWA solver, and arrived at a dimerized grating (period 962.5~nm, height 350~nm, two 235~nm wide silicon bars, one displaced by 24~nm) with $\lambda_0 = 1550.16$~nm and $Q = 1025$ at the converged truncation. FDTD with harmonic inversion gave $\lambda_0 = 1549.49$~nm and $Q = 971$.
The verification again exposed a numerical artifact, this time on the FDTD side: at resolutions for which the period or half-period does not span an integer number of grid cells, the discretized cell is no longer symmetric, and $Q$ took spurious values of 1292 and 1908, whereas commensurate grids gave 998 and 971. In the same session, a reflection quasi-BIC in Si at 532~nm was abandoned after the convergence check showed that the narrow peaks found at low truncation were largely artifacts, and the converged spectra confirmed that absorption limits $Q$ to a few tens in this material.

Over the work reported here, excluding the session ~S11 of Supplementary Material, the agent executed 84 bridge requests and 134 cluster jobs, including 96 optimizations (11.2 GPU-hours) and 14 FDTD validation jobs. It detected eight issues and proposed a corrective action for each (sec.~S9). Four of them would have led to wrong conclusions had only the nominal efficiency computed by the optimization model been examined: the most efficient design would have been chosen although it is the most sensitive to fabrication errors; the threshold-based robust formulation would have been accepted although it imposed edge shifts three to four times larger than intended; meta-atoms trapped in local optima would have been kept; and the y-polarized efficiency of the metagrating would have been overestimated by 5\%.

The agent does not know more physics than an experienced designer: every check it performs is standard. The difference is that the checks are applied to every run, by a fixed and documented protocol, and that a second rigorous method decides whether a result is reported. This approach is complementary to agents based on neural surrogates \cite{Lupoiu2025}. A surrogate can propose candidate designs quickly; the proposed protocol can then optimize them with a rigorous solver, make them robust to fabrication errors and verify them with an independent solver (see sec.~S10 of Supplementary Material).

The work has four main limitations. First, the robust formulation does not enforce a minimum feature size: narrow features are reported but not prevented, which would require an explicit length-scale constraint \cite{Hammond2021}. Second, the metalenses are validated, but not optimized, with full-wave simulations; large-area optimization methods \cite{Pestourie2018, Hammond2022} could recover part of the efficiency lost to coupling between neighboring atoms. Third, beyond one additional specification addressed in a new session (see sec.~S11 of Supplementary Material), we did not evaluate the agent systematically across different specifications or language models. Finally, all results are numerical: their reliability rests on convergence within each solver and on the agreement between two independent methods, and the released layouts are intended for experimental verification.

In summary, coupling an AI agent to a differentiable 3D solver with exact fabrication robustness and a mandatory independent full-wave verification yields designs whose reported figures can be trusted. In the cases presented here, the independent verification detected numerical artifacts in both solvers: a polarization-dependent truncation error in RCWA that had passed a standard convergence test, and a grid-induced symmetry breaking in FDTD that distorted the quality factor of a quasi-BIC. As agents take over more of the simulation loop, cross-validation between independent methods should become part of the loop rather than an afterthought.

\section*{Supplementary Material}
See the supplementary material for the execution bridge and security model, the optimization and robustness
formulations, the analysis checklist and validation protocol, the full design histories, the meta-atom library,
supercells and refinement, the metalens model, field-level comparisons, the process record and a comparison with
other AI-driven design frameworks.

\begin{acknowledgments}
G.E.L. thanks the CINECA projects DENZMODE (ID:HP10CILT35)  under the ISCRA initiative, for the availability of high-performance computing resources and support. During this work the authors used an AI coding agent (Claude, Anthropic) to write and operate the software, run and analyse the simulations. The authors reviewed and edited all content and take full responsibility for it.
\end{acknowledgments}

\section*{Author Declarations}
\subsection*{Conflict of Interest}
The authors have no conflicts to disclose.

\section*{Data Availability}
The code, the agent instructions and all runs discussed here are openly available on GitHub at
\url{https://github.com/GiuseppeELio/metasurface-agent-pipeline} (MIT License) and archived on Zenodo at 10.5281/zenodo.23982523. TORCWA and Meep are installed separately under their own licences.
\clearpage
\bibliography{refs}
\end{document}


\title{Supplementary Material\\[4pt]\large An Agent-Driven, Cross-Validated Inverse-Design Pipeline for Dielectric Metasurfaces}

\author[1]{Giuseppe Emanuele Lio\thanks{giuseppe.lio@unipi.it}}
\author[2,3]{Roberto Caputo}
\author[4]{Alain J. Corso}
\author[4,5]{Maria G. Pelizzo}
\author[3]{Antonio Ferraro}
\affil[1]{Department of Physics ``E. Fermi'', University of Pisa, Largo B. Pontecorvo 3, 56127 Pisa, Italy}
\affil[2]{Department of Physics, University of Calabria, 87036 Rende (CS), Italy}
\affil[3]{Consiglio Nazionale delle Ricerche - Istituto di Nanotecnologia (CNR-Nanotec), 87036 Rende (CS), Italy}
\affil[4]{Consiglio Nazionale delle Ricerche - Istituto di Fotonica e Nanotecnologie (CNR-IFN), Via Trasea 7, 35131 Padova, Italy}
\affil[5]{Dipartimento di Ingegneria dell'Informazione, Universit\`a degli Studi di Padova, Via G. Gradenigo 6/B, 35131 Padova, Italy}

\date{}
\maketitle
\tableofcontents

\section{Agent workflow, execution bridge and security model}
\label{sec:bridge}
Figure~1 of the main text summarizes the loop. The user states a target in plain language, for instance
\emph{``maximize the +1 transmitted order at 1550~nm for unpolarized light, silicon on silicon, air above, robust to
$\pm$10~nm''}. The agent checks feasibility (propagating orders, phase range versus thickness, index contrast),
writes a configuration file and, when several runs are needed, a job list; it estimates the cost and asks before
spending significant compute. Jobs are executed by the bridge on the target machine. When results return, the agent
applies the checklist of Section~\ref{sec:checklist}, reports a short table (nominal and worst-case efficiency per
polarization, dominant loss channel), gives a physical interpretation and proposes exactly one next step with its
estimated cost. All instructions for the agent are kept in a single plain-text file in the repository
(\texttt{AGENTS.md}), so that the behaviour is transparent, versioned and reproducible with any capable coding agent.

The agent never connects to a cluster and never sees passwords, one-time codes or keys. A request is a text file
\texttt{queue/<name>.req} containing a single line with one of four allowed sub-commands: \texttt{submit} (one
optimization), \texttt{array} (a list of runs or post-processing scripts), \texttt{meep} (a list of FDTD
validations on a CPU node) and \texttt{check} (a short GPU test). A watcher started by the user in their own terminal
validates the line against this whitelist and rejects shell metacharacters, synchronizes the code, submits the job,
follows it and retrieves the results; its state is exposed through small status files (heartbeat, queue snapshot,
done and failed folders). Authentication, including two-factor login, is performed by the user. All site-specific
information (host, account, partitions, environment modules) lives in a single git-ignored file that the agent fills
in with the user during a first-run onboarding; in its absence the same protocol runs jobs on the local machine.
Large intermediate files (checkpoints) stay on the target and can be reused for warm starts.

\section{Electromagnetic model, optimization and fabrication robustness}
\label{sec:em}
A single patterned layer of thickness $h$ on a substrate is described by a density $\rho(x,y) \in [0,1]$ on a
pixel grid of 5~nm, mapped to the permittivity $\varepsilon = \rho\,\varepsilon_\mathrm{core} +
(1-\rho)\,\varepsilon_\mathrm{bg}$. The scattering matrix is computed by TORCWA~\cite{Kim2023}, a PyTorch
implementation of RCWA with automatic differentiation, in single precision on the GPU. For a figure of merit (FoM)
$F(\rho)$, the gradient $\partial F / \partial \rho$ is obtained by reverse-mode differentiation through the
eigendecomposition and is used by the Adam optimizer~\cite{Kingma2015}.

Two objectives are used in this work. For deflectors,
\begin{equation}
F_\mathrm{defl} = \frac{1}{2} \sum_{p \in \{x,y\}} \eta^{(p)}_{+1},
\qquad \eta^{(p)}_{m} = \sum_{q \in \{x,y\}} \left| S^{(qp)}_{m} \right|^2,
\label{eq:fdefl}
\end{equation}
where $S^{(qp)}_m$ is the power-normalized transmission amplitude into order $m$ and output polarization $q$ for
input polarization $p$; the average over $p$ corresponds to unpolarized light. For meta-atoms with a target phase
$\phi_\mathrm{t}$,
\begin{equation}
F_\mathrm{atom} = \frac{1}{2} \sum_{p} \mathrm{Re}\!\left[ t^{(p)}_{00}\, e^{-i(\phi_\mathrm{t} + \phi_\mathrm{ref})} \right],
\label{eq:fatom}
\end{equation}
which rewards both a high transmission $|t_{00}|^2$ and a small phase error; $\phi_\mathrm{ref}$ is the phase of
the unpatterned cell of equal thickness. The deflection angle of order $m$ follows the grating equation
$\sin\theta_m = m\lambda / (n_\mathrm{out}\Lambda_x)$.

The density is regularized by a periodic Gaussian filter of radius $R$ (40~nm here),
$\tilde\rho = G_R * \rho$, and projected with
\begin{equation}
\bar\rho = \frac{\tanh(\beta\eta) + \tanh(\beta(\tilde\rho - \eta))}{\tanh(\beta\eta) + \tanh(\beta(1-\eta))},
\label{eq:projection}
\end{equation}
with $\eta = 0.5$ and a geometric continuation of $\beta$ up to $10^3$, which drives the design to a binary
pattern~\cite{Wang2011}. Mirror symmetry in $x$, $y$ or both, or the full $D_4$ symmetry of the square cell, is
imposed by symmetrizing $\rho$; $D_4$ makes the zero-order response independent of polarization. Initial
geometries (disk, square, ring, cross, four-leaf clover) can be specified by their size, and any previous design or
checkpoint can be used as a warm start.

\label{sec:robust}
Edge-placement errors of a few nanometres are typical of electron-beam lithography and etching. We use a geometric
robust formulation in which, at every iteration, the projected design is re-filtered with a second Gaussian kernel
of radius $R_2 = 25$~nm and thresholded at $t_{e,d} = \tfrac{1}{2}\left[1 \pm \mathrm{erf}(d/R_2)\right]$,
which for a straight boundary moves the edge inwards (eroded) or outwards (dilated) by exactly $d = 10$~nm. The
three designs are simulated and combined with a smooth minimum
$F_\mathrm{rob} = -k^{-1} \ln \sum_{j} e^{-k F_j}$ ($k = 10$). This differs from the classical threshold-based
formulation~\cite{Wang2011}, in which the physical edge shift depends on the local grey level of the filtered density.
In our first robust runs the threshold variant corresponded to effective shifts of 17--20~nm instead of the intended
5~nm, which made the optimization overly conservative. The geometric variant removes this dependence, so that the
tolerance specified by the user is the tolerance actually enforced, independently of the grey level of the design
during the optimization. Robustness must be active
from the first iteration: activating it late, on an almost binary design, stalls the optimizer.

Independently of the training formulation, every final binary design is evaluated with \emph{exact} edge shifts of
$\pm$5, $\pm$10 and $\pm$15~nm computed with periodic Euclidean distance transforms, and with a minimum-feature
report (fraction of solid and void narrower than 30, 40, 60 and 80~nm, from morphological openings). Designs are
ranked by the worst case within the expected tolerance, not by the nominal FoM.

\section{Analysis checklist}
\label{sec:checklist}
Each run writes a summary with the FoM of the continuous and binary designs, the efficiency and angle of every
propagating order in transmission and reflection for both input polarizations, the energy balance, a convergence
check at a higher Fourier truncation, the robustness and feature-size reports and the grey level. The agent applies
the same checklist to every run: (1) energy balance, $T + R = 1$ within 1\% for lossless materials; (2) convergence,
with a relative change below 2--3\% at the verification truncation; (3) binarization; (4) worst-case robustness;
(5) minimum features; (6) where the lost power goes (parasitic orders or reflection, per polarization), which
dictates the next change; (7) local optima, checked by other seeds and initial shapes; (8) plots in physical units,
with field-derived efficiencies cross-checked against the scattering parameters; and (9) the cross-solver verdict
(Section~\ref{sec:validation}) before a design is reported as final.

\section{Cross-solver validation protocol and results}
\label{sec:validation}
RCWA and FDTD make different numerical errors: RCWA suffers from Fourier truncation and from the slow convergence of
field discontinuities at high index contrast~\cite{Li1996}, whereas FDTD suffers from staircasing or subpixel
smoothing, the rounding of the cell to whole pixels and imperfect absorbing layers. Agreement between them is
therefore strong evidence that a result is correct. The pipeline keeps both solvers: TORCWA for optimization, and
Meep~\cite{Oskooi2010}, running with MPI on CPU nodes, as an independent referee.

For a periodic design, Meep simulates one cell with Bloch-periodic lateral boundaries, perfectly matched layers along
$z$, and a pulsed plane-wave source in the input medium. The efficiency of every propagating transmitted order is
obtained by projecting the transmitted fields on the s- and p-polarized diffracted plane waves of that order, and
normalized by the incident power computed in a reference run filled with the input medium. The binary design enters
as a material grid with bilinear interpolation; the Courant factor is reduced to 0.35, because the default value
together with gradient-based subpixel averaging of high-index binary grids produced numerical instabilities at
resolutions of 50--60~pixels/$\upmu$m.

A validation round is planned with one command. It produces a GPU job list (RCWA references at the verification
truncation, nominal and edge-shifted, plus near fields) and an FDTD job list (each requested resolution and edge
shift), together with a run-time estimate. After both have run, a report compares the target-order efficiency
$\eta$ and the total transmission $T$ of the two solvers and assigns a verdict per design, edge shift and
polarization: \emph{agree} if $|\Delta\eta| \le 0.03$ and $|\Delta T| \le 0.02$; \emph{inconclusive} if not, and
only one FDTD resolution is available or FDTD still changes by more than half the gap between its two highest
resolutions; \emph{marginal} if the gap is within twice the tolerance; \emph{disagree} otherwise. For a
\emph{disagree} verdict the agent must diagnose before reporting: raise the RCWA truncation, raise the FDTD
resolution, and check that both solvers see the same geometry. This ladder turns a disagreement into a diagnosis
rather than a mere flag: in this work it traced a polarization-dependent gap to the slow convergence of the RCWA
expansion for the field component normal to the high-contrast boundaries~\cite{Li1996}, i.e., to the optimization
model and not to the design (Section~\ref{sec:validres}). Between design stages the ranking of candidates
matters more than absolute values, so the report also gives the fraction of candidate pairs ordered identically by
the two solvers. The user chooses, once, whether validation runs only on final designs (default), at every decision
point (quick, single-resolution checks) or not at all.

\label{sec:validres}

\begin{figure}[t]
\centering
\includegraphics[width=0.85\linewidth]{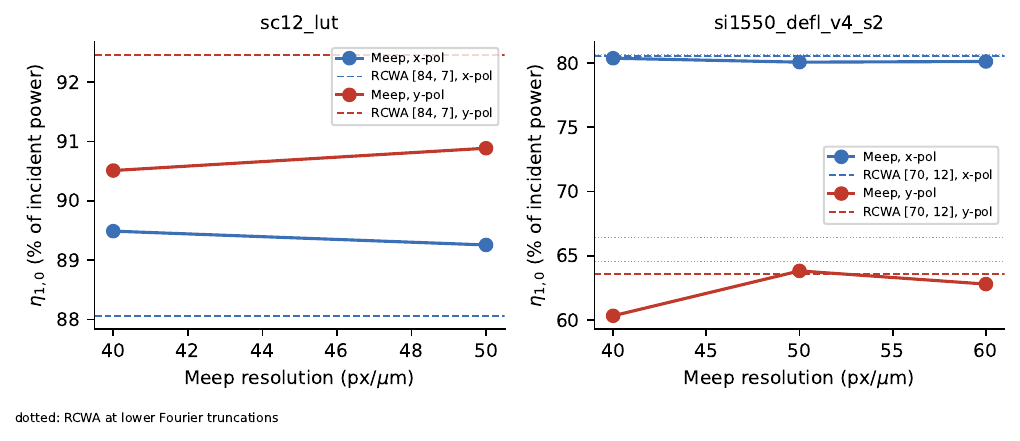}
\caption{Resolution study of the FDTD validation. +1-order efficiency from Meep (markers) versus resolution, for the
12-atom supercell (left) and the robust metagrating v4 s2 (right), and RCWA references (dashed: highest truncation;
dotted: lower truncations). For the metagrating in y polarization, the RCWA value converges towards the FDTD value
as the truncation increases.}
\label{fig:meep}
\end{figure}

\begin{figure}[t]
\centering
\includegraphics[width=0.49\linewidth]{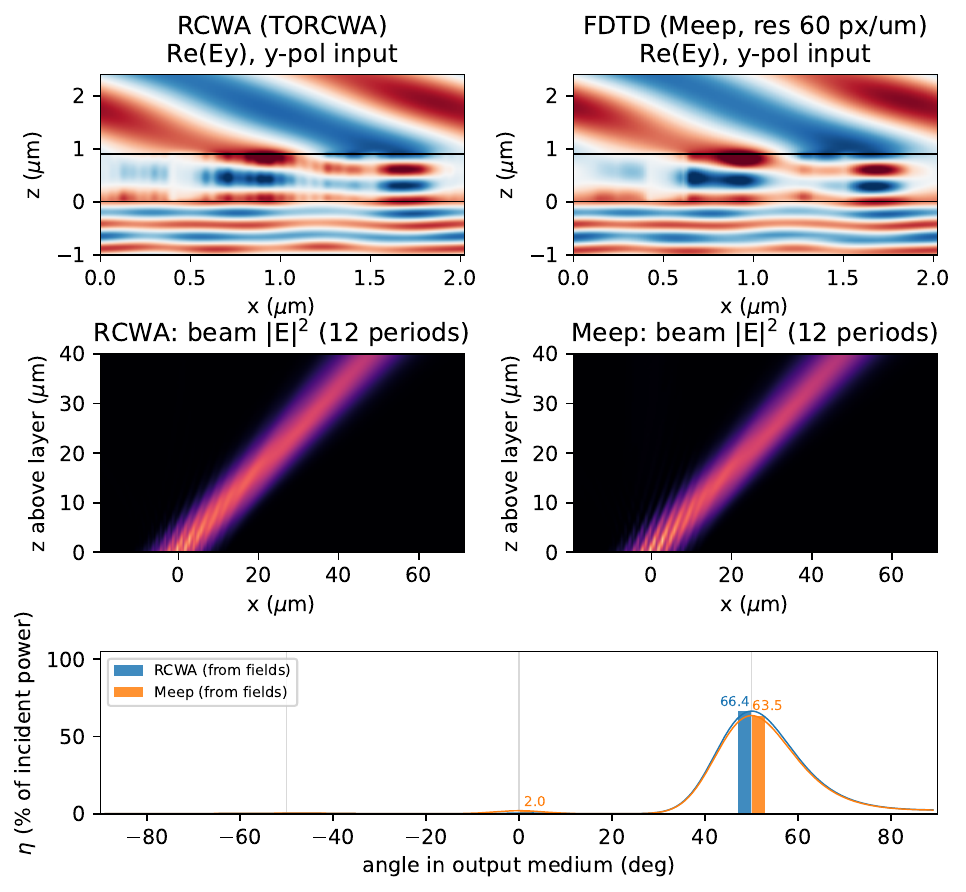}\hfill
\includegraphics[width=0.49\linewidth]{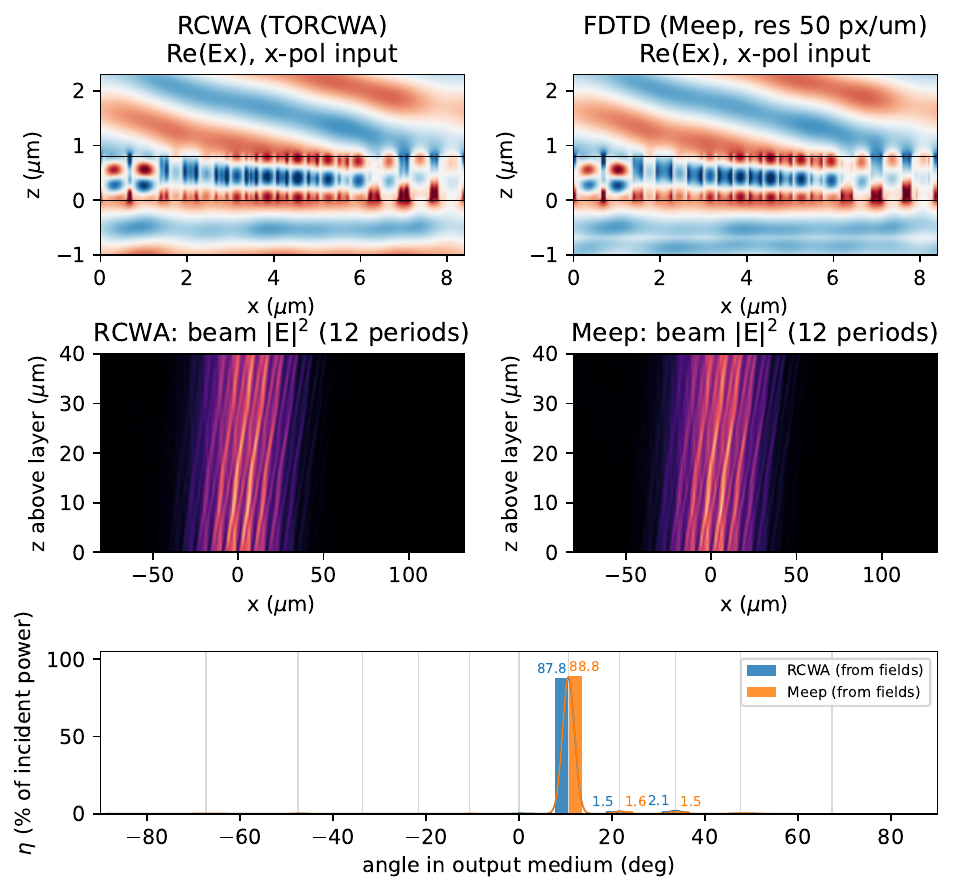}
\caption{Field-level comparison of TORCWA and Meep for (\textbf{a}) the metagrating v4 s2, y-polarized input, and
(\textbf{b}) the 12-atom supercell, x-polarized input. Top: real part of the incident-polarization field component in
the $xz$ plane (the horizontal lines mark the layer). Middle: intensity of a beam formed by a Gaussian-apodized
aperture of 12 periods, propagated in air with the angular-spectrum method from the near field above the layer.
Bottom: efficiencies of the transmitted orders in percent of the incident power, computed from the Fourier
coefficients of all field components.}
\label{fig:fields}
\end{figure}

The robust metagrating v4~s2 and the 12-atom library supercell were validated as final designs
(Table~\ref{tab:valid}, Figures~\ref{fig:meep} and \ref{fig:fields}). For x polarization the two solvers agree to
within 0.5 points over the whole resolution study. For y polarization, the first comparison, against RCWA at the
verification truncation $[35,7]$, gave a \emph{marginal} verdict ($\eta = 0.664$ versus 0.628). Following the
diagnosis prescribed by the protocol, the RCWA truncation was raised: the RCWA value decreases monotonically
(0.664, 0.646, 0.636 at $[35,7]$, $[50,10]$, $[70,12]$) towards the FDTD value, which itself changes by only
1 point between 50 and 60~pixels/$\upmu$m, and the verdict becomes \emph{agree}. The error was therefore in the
model used for optimization, and it affected only one polarization, consistent with the known slow convergence of
RCWA for field components normal to high-contrast boundaries~\cite{Li1996}. For the supercell the
solvers agree within 1.6 points for both polarizations, and both rank the two designs identically. Near fields
and the far-field order spectra computed from them (Figure~\ref{fig:fields}) overlap; the efficiencies derived from
the fields reproduce the scattering-matrix values. Each FDTD run takes about 1~min (metagrating) to 25~min
(supercell) at 40--60~pixels/$\upmu$m.

\begin{table}[t]
\caption{Cross-solver validation of the final periodic designs: efficiency $\eta$ of the +1 transmitted order and
total transmission $T$ from RCWA (TORCWA) at the highest truncation and from FDTD (Meep) at the highest resolution.}
\label{tab:valid}
\centering
\small
\begin{tabular}{llccccl}
\toprule
Design & Pol. & $\eta$ RCWA & $\eta$ FDTD & $T$ RCWA & $T$ FDTD & Verdict \\
\midrule
Metagrating v4 s2 ($[70,12]$; 60 px/$\upmu$m) & x & 0.806 & 0.801 & 0.818 & 0.814 & agree \\
                                              & y & 0.636 & 0.628 & 0.657 & 0.650 & agree \\
Supercell N = 12 ($[84,7]$; 50 px/$\upmu$m)    & x & 0.881 & 0.892 & 0.933 & 0.940 & agree \\
                                              & y & 0.925 & 0.909 & 0.938 & 0.922 & agree \\
\bottomrule
\end{tabular}
\end{table}

\section{Case 1: full design history of the metagrating}
\label{sec:case1}
The first specification was a transmissive grating on a silicon substrate, with air above, that sends unpolarized
light at normal incidence into the +1 order at 50$^\circ$. The grating equation gives $\Lambda_x = 2023$~nm; a
period $\Lambda_y = 400$~nm keeps the response one-dimensional in terms of diffraction. The cell is discretized on a
$405 \times 80$ grid, with symmetry in $y$, $[25,5]$ Fourier harmonics during optimization and $[35,7]$ for
verification; one forward-plus-backward evaluation takes about 1.3~s. Table~\ref{tab:case1} and Fig.~2 of the main text
summarize the progression of design rounds proposed by the agent.

\begin{table}[t]
\caption{Case~1: +1-order efficiency for unpolarized light of selected binary designs, nominal and worst case under
exact edge shifts, evaluated at the $[35,7]$ truncation (v1 at $[30,6]$). Last column: fraction of the solid in
features narrower than 40~nm.}
\label{tab:case1}
\centering
\small
\begin{tabular}{llcccccc}
\toprule
Design & Description & $h$ (nm) & Nominal & $\pm$5~nm & $\pm$10~nm & $\pm$15~nm & Solid $<$40~nm \\
\midrule
v1     & first run, no robustness           & 700  & 0.736 & 0.695 & 0.652 & 0.597 & 0.24\% \\
v2 h600 & thickness sweep                   & 600  & 0.778 & 0.737 & 0.676 & 0.592 & 0.03\% \\
v2 h800 & thickness sweep                   & 800  & 0.778 & 0.742 & 0.678 & 0.583 & 0.24\% \\
v2 h900 & thickness sweep, best nominal     & 900  & \textbf{0.810} & 0.667 & 0.607 & 0.510 & 0.60\% \\
v3 s1  & threshold robustness              & 900  & 0.692 & 0.671 & 0.627 & 0.568 & 0.02\% \\
v4 s1  & geometric robustness $\pm$10~nm   & 900  & 0.674 & 0.674 & 0.674 & 0.639 & 3.2\% \\
v4 s2  & geometric robustness $\pm$10~nm   & 900  & 0.735 & 0.681 & \textbf{0.681} & \textbf{0.680} & 4.0\% \\
v4 s3  & geometric robustness $\pm$10~nm   & 900  & 0.699 & 0.670 & 0.670 & 0.623 & 0.08\% \\
\bottomrule
\end{tabular}
\end{table}

The first run (v1) reached 0.74. The agent then proposed a thickness sweep (v2, 600--1100~nm, one run per
thickness), which produced the best nominal design, $h = 900$~nm with 0.81. The robustness report showed that this
design is also among the most fragile: its efficiency falls to 0.61 for $\pm$10~nm and to 0.51 for $\pm$15~nm edge
shifts (Fig.~2(b) of the main text). A first robust round (v3) with threshold-based erosion and dilation was overly
conservative (Section~\ref{sec:robust}); the geometric formulation (v4, three seeds) produced v4~s2, whose response
is almost flat over $\pm$15~nm and whose worst case within $\pm$10~nm, 0.68, is the highest of all runs (the non-robust v2 h800 comes close, 0.678, but degrades faster at $\pm$15~nm). The price
is 7.5 points of nominal efficiency. Fig.~2(c) of the main text compares the two patterns: the robust design
avoids the narrow bridges and gaps of the nominal optimum, although 4\% of its solid area is still in features below
40~nm, because a geometric robust formulation does not by itself enforce a minimum length scale~\cite{Hammond2021}.

The per-order analysis identifies the remaining loss channel: for v4~s2 the +1 efficiency is 0.81 for x- and 0.68 for
y-polarized input, and the y polarization loses 32\% of the power to reflection into the high-index substrate.
An anti-reflection layer or a bilayer design is the natural next step for this specification.

\section{Meta-atom library}
\label{sec:library}

\begin{figure}[t]
\centering
\includegraphics[width=0.98\linewidth]{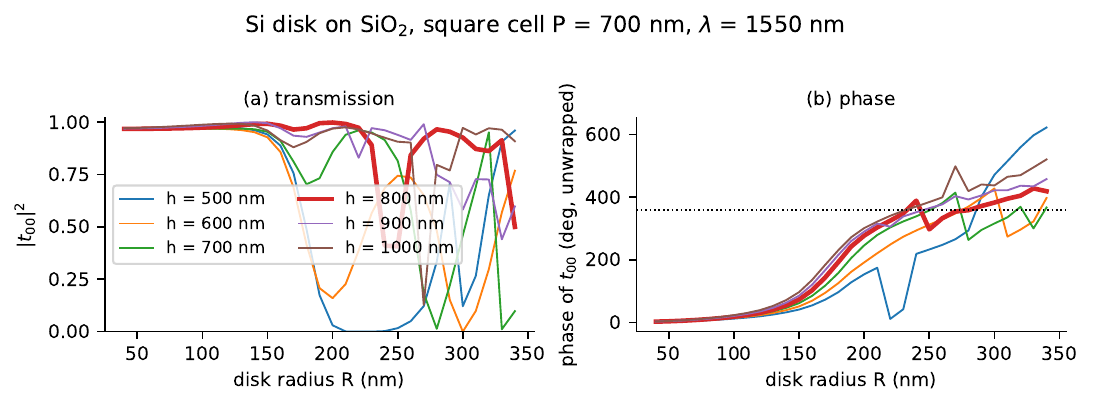}
\caption{Parametric pre-screening: (\textbf{a}) zero-order transmission and (\textbf{b}) unwrapped phase of silicon disks
on SiO$_2$ versus radius, for thicknesses of 500--1000~nm ($P = 700$~nm, $\lambda = 1550$~nm). The dotted line marks
360$^\circ$. The selected thickness, $h = 800$~nm, is highlighted.}
\label{fig:disk}
\end{figure}

The second specification asked for square-cell silicon meta-atoms on glass, obtained from a disk by freeform erosion
into symmetric shapes, with maximum transmission and a set of phase levels, to be assembled into supercells that
deflect unpolarized light into the +1 order. The user initially proposed a thickness of 100~nm; the agent's
feasibility check showed that a 100~nm disk covers only about 50$^\circ$ of phase, so the thickness was agreed after
a parametric sweep. Figure~\ref{fig:disk} shows the transmission and phase of silicon disks on SiO$_2$
($n = 1.444$) in a $P = 700$~nm square lattice: $h = 800$~nm is the smallest thickness that covers more than
360$^\circ$ with high transmission, with a steep phase variation between $R = 150$ and 200~nm and a transmission
dip near $R = 250$~nm.

\begin{figure}[t]
\centering
\includegraphics[width=0.98\linewidth]{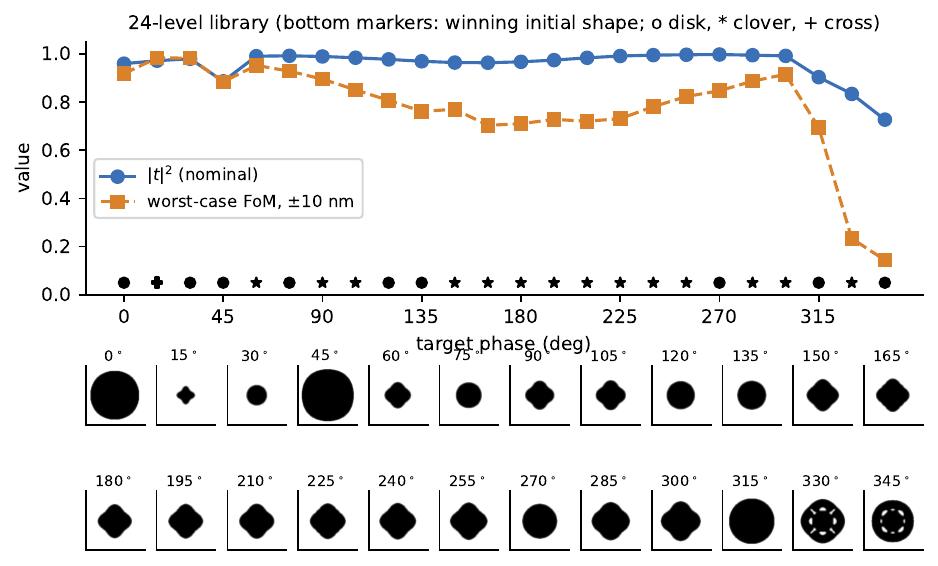}
\caption{The 24-level meta-atom library ($h = 800$~nm Si on SiO$_2$, $P = 700$~nm, $D_4$ symmetry). Top: nominal
transmission $|t|^2$ (blue) and worst-case FoM of Eq.~(\ref{eq:fatom}) under $\pm$10~nm edge shifts (orange) versus
target phase; the markers at the bottom indicate which initial shape won (disk, clover or cross). Bottom: the
selected unit cells, 700~nm wide.}
\label{fig:lut}
\end{figure}

Each meta-atom was optimized with the objective of Eq.~(\ref{eq:fatom}), $D_4$ symmetry, geometric robustness of
$\pm$10~nm, $[9,9]$ harmonics ($[13,13]$ for verification) and 200 iterations (about 5~min per atom). A first attempt
in which all phase levels started from the same $R = 300$~nm disk failed: only three of eight phases were reached,
the others being trapped in local optima, and the resulting supercell reached only 0.23 efficiency. Following the
checklist item on local optima, the agent proposed physics-informed initializations: for each target phase, a disk
with the radius that already provides that phase in the sweep of Figure~\ref{fig:disk}, plus a clover and a cross
of equal area. The final library has 24 levels in steps of 15$^\circ$ (48 additional optimizations, about 4.5~h on one GPU, complementing the eight levels of the earlier attempt);
for every level the atom with the best worst-case FoM within $\pm$10~nm, for x and y polarization together, was
selected (Figure~\ref{fig:lut}).

The nominal transmission exceeds 0.95 for 20 of the 24 levels. Robustness varies strongly across the library.
Small atoms are very robust (worst-case FoM of 0.93--0.98 for 15, 30, 60 and 75$^\circ$); the intermediate phases (165--225$^\circ$) are
the fabrication bottleneck at about 0.70--0.73; the levels at 330$^\circ$ and 345$^\circ$, which require radii near
the transmission dip of Figure~\ref{fig:disk}, are fragile (0.23 and 0.14). The clover shape wins 13 of the 24 levels,
showing that freeform modification of the disk adds useful degrees of freedom.

\section{Supercells and global robust refinement}
\label{sec:supercell}

\begin{figure}[t]
\centering
\includegraphics[width=0.98\linewidth]{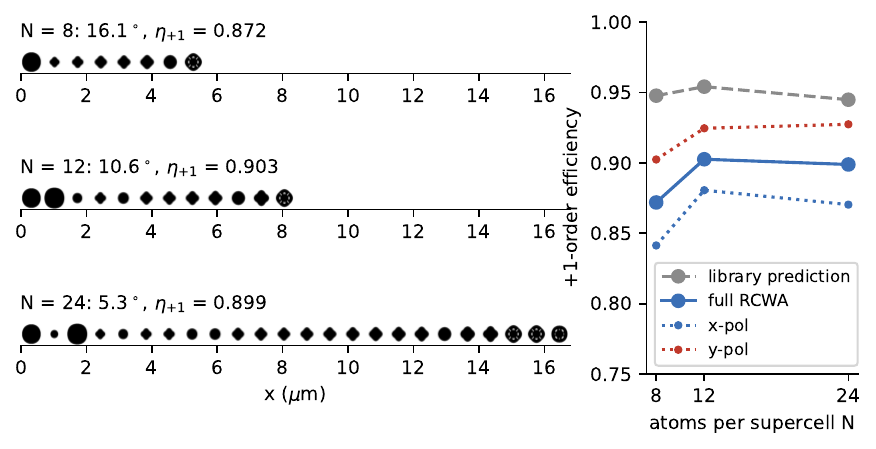}
\caption{Supercells assembled from the library. Left: layouts at true scale for $N = 8$, 12 and 24 atoms, with the
deflection angle and the full-RCWA +1 efficiency for unpolarized light. Right: efficiency versus $N$, from the library
(independent-atom) prediction and from full-supercell RCWA, with the x- and y-polarized contributions.}
\label{fig:supercells}
\end{figure}

A supercell of $N$ atoms, with phases closest to $k\,360^\circ/N$, deflects into the +1 order at
$\sin\theta = \lambda/(N P)$. Table~\ref{tab:sc} and Figure~\ref{fig:supercells} compare the full-supercell RCWA
efficiency with the \emph{library prediction}, the +1 Fourier coefficient of the sequence of atom transmissions
$\sqrt{T_k}\,e^{i\phi_k}$, which assumes independent atoms. All three supercells exceed 0.87 for unpolarized light.
The gap with the library prediction, which measures near-field coupling between neighbouring atoms, decreases from
7.6 points at $N = 8$ to 5.1 and 4.6 points at $N = 12$ and 24, as the phase gradient becomes smoother.

\begin{table}[t]
\caption{Supercells assembled from the library, and their global robust refinement. Efficiency of the +1 transmitted
order for unpolarized light (x / y input in parentheses), from full-supercell RCWA.}
\label{tab:sc}
\centering
\small
\begin{tabular}{lcccccc}
\toprule
Supercell & $\theta$ & Library prediction & Nominal & $-$10~nm & $+$10~nm & Worst case \\
\midrule
N = 8           & 16.07$^\circ$ & 0.948 & 0.872 (0.841 / 0.902) & 0.816 & 0.701 & 0.701 \\
N = 8, refined  & 16.07$^\circ$ & --    & 0.782 (0.851 / 0.713) & 0.779 & 0.815 & 0.779 \\
N = 12          & 10.63$^\circ$ & 0.954 & 0.903 (0.881 / 0.925) & 0.879 & 0.749 & 0.749 \\
N = 12, refined & 10.63$^\circ$ & --    & 0.889 (0.900 / 0.878) & 0.848 & 0.850 & \textbf{0.848} \\
N = 24          & 5.29$^\circ$  & 0.945 & 0.899 (0.870 / 0.927) & --    & --    & -- \\
\bottomrule
\end{tabular}
\end{table}

\begin{figure}[t]
\centering
\includegraphics[width=0.98\linewidth]{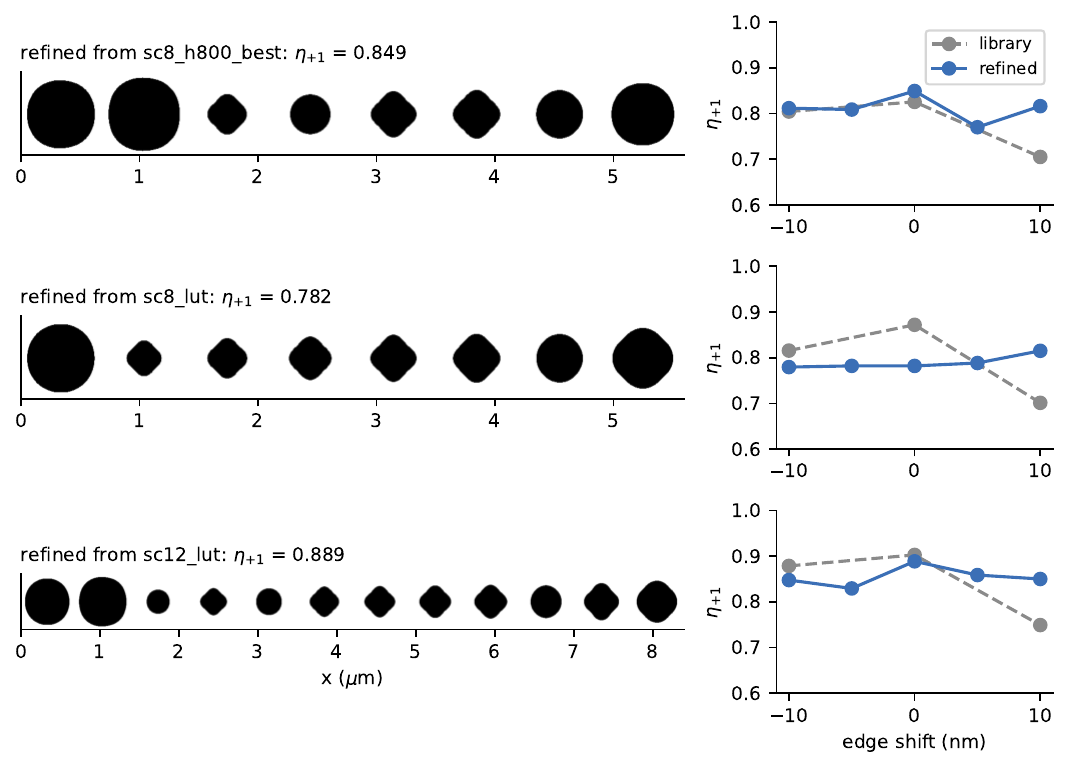}
\caption{Global robust refinement of assembled supercells. Left: refined layouts. Right: +1 efficiency versus exact
edge shift for the starting library supercell (grey) and the refined design (blue). From top to bottom: an early
8-atom supercell, the 8-atom and the 12-atom supercells of Table~\ref{tab:sc}.}
\label{fig:refine}
\end{figure}

All library supercells are most sensitive to dilation: at $+$10~nm the largest atoms, around 0$^\circ$ and
342$^\circ$, nearly touch their neighbours. The assembled supercell is therefore a natural warm start for a
\emph{global} robust topology optimization of the whole supercell, with the same filters and $\pm$10~nm geometric
robustness as above, $[48,5]$ harmonics for $N = 12$, $\beta$ starting at 30 and 120 iterations (50~min).
For $N = 12$ the refinement trades 1.4 points of nominal efficiency for 10 points of worst case (0.749 to 0.848),
and balances the two polarizations (Figure~\ref{fig:refine}). For $N = 8$ the worst case also improves (0.70 to
0.78), but the nominal efficiency drops by 9 points: the Gaussian filter cannot represent the finest features of the
library atoms, so the warm start itself is degraded (0.76 at the first iteration). This suggests matching the filter
radius of the refinement to the feature size of the library.

\section{Metalens model and full-wave comparison}
\label{sec:lens}

The same library also defines non-periodic devices. For a circular metalens of diameter $D$ and focal length $f$,
the target phase is $\phi(r) = -k_0 n_\mathrm{out}\left(\sqrt{r^2 + f^2} - f\right)$, sampled at the centre of each
cell; each cell receives the atom $k$ that maximizes $\sqrt{T_k}\cos(\phi_k - \phi)$, which accounts for both phase
error and transmission, and a global phase offset is chosen to maximize the total projection. A fast vector model
assumes that each cell transmits as in its periodic lattice (locally periodic approximation~\cite{Pestourie2018}) and
propagates the transmitted field with the angular-spectrum method, keeping all propagating plane waves and deriving
the longitudinal field component and the magnetic field from transversality. Because only propagating components are
stored, the full analysis takes a few seconds. It gives the intensity in the $xz$ and $yz$ planes, from the lens
surface through the focus; the on-axis profile, whose maximum defines the focus and whose full width at half maximum
defines the depth of focus (DOF); the lateral FWHM in $x$ and $y$ at the focus; the focusing efficiency (power through
a disk of radius three times the lateral FWHM, divided by the incident power on the aperture); and the Strehl ratio
relative to an ideal lens of the same aperture.

The full-wave check simulates the whole aperture in three dimensions with Meep. The focal field is obtained in two
independent ways: with Meep's near-to-far-field transformation, and by recording the transverse field 0.4~$\upmu$m
above the lens and propagating it with the \emph{same} angular-spectrum code as the model. In the second route the
model and the FDTD results share the propagator and every definition, so that any difference comes from the near
field alone; the two routes agree within 0.4 points of focusing efficiency and 0.6~$\upmu$m in focal position.

\begin{figure}[t]
\centering
\includegraphics[width=0.98\linewidth]{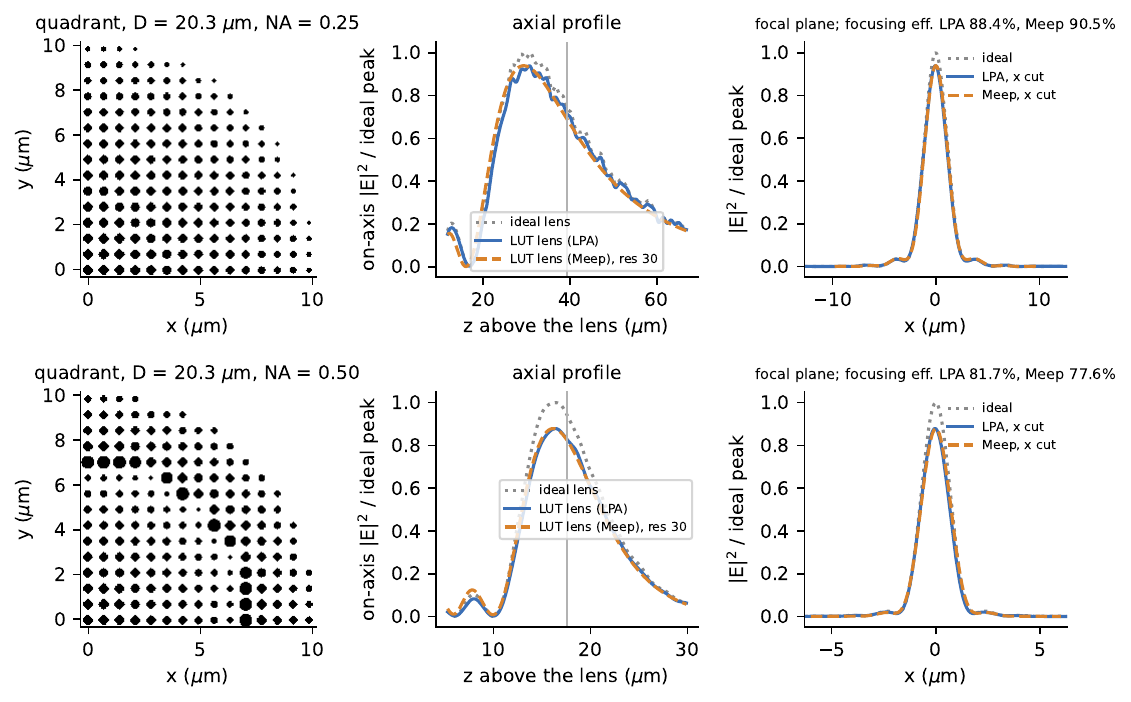}
\caption{Metalenses built from the 24-level library, NA 0.25 (top) and NA 0.5 (bottom), $D = 20.3\um$. Left: one
quadrant of the layout (the optical axis is at the origin). Centre: on-axis intensity versus distance from the lens
for the ideal lens, the locally periodic model (LPA) and FDTD (Meep, 30~pixels/$\upmu$m, scaled to the LPA peak
for shape comparison); the vertical line marks the design focal length. Right: focal-plane intensity cut along $x$;
the title gives the focusing efficiencies (FDTD values from the near-to-far transformation; Table~\ref{tab:lens}
reports the angular-spectrum route, which agrees within 0.4 points).}
\label{fig:lens}
\end{figure}

\begin{figure}[t]
\centering
\includegraphics[width=0.9\linewidth]{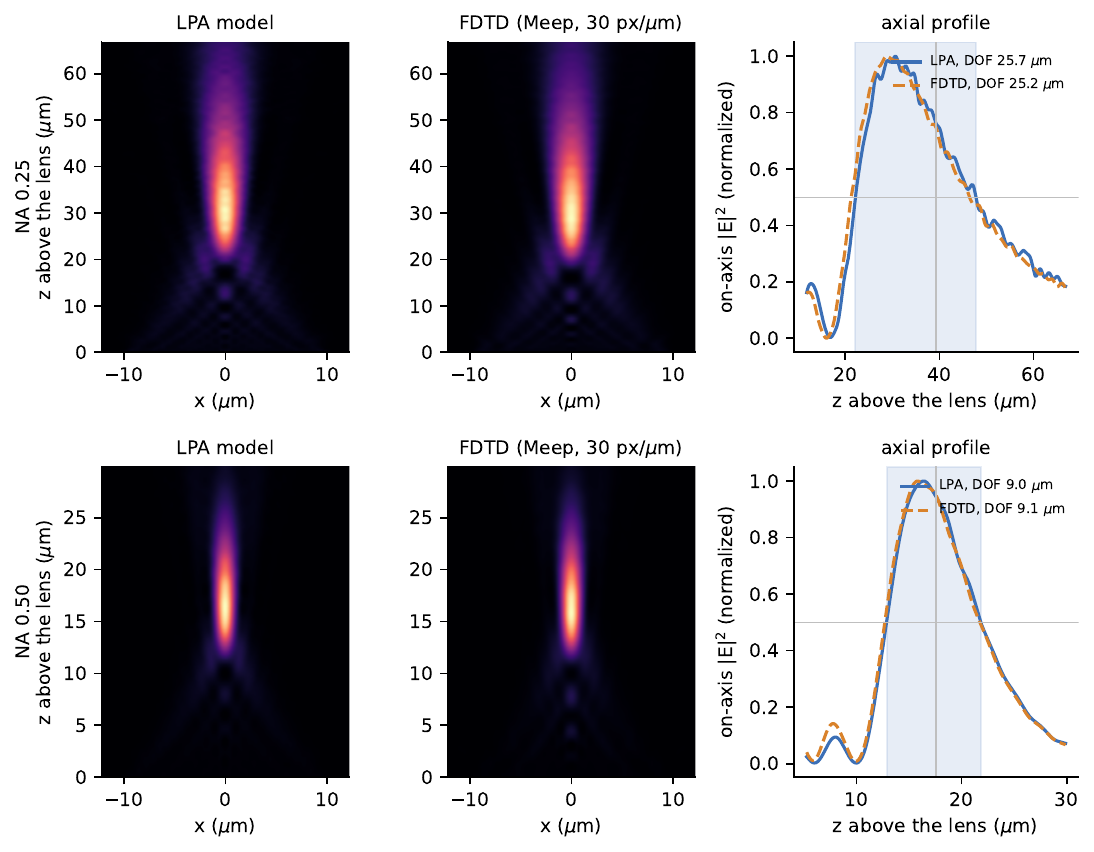}
\caption{Propagation behind the metalenses, NA 0.25 (top) and NA 0.5 (bottom), under normal illumination from the
substrate. Left and centre: intensity $|E|^2$ in the $xz$ plane through the optical axis, from the lens surface, for the
locally periodic model and for the FDTD near field propagated with the same angular-spectrum code (each normalized to
its maximum). Right: on-axis intensity; the shaded band is the depth of focus of the model, and the vertical line marks
the design focal length. The NA 0.5 case is also shown in Fig.~4 of the main text.}
\label{fig:lensprop}
\end{figure}

Two lenses of $D = 20.3\um$ ($29 \times 29$ atoms) were designed, with NA 0.5 and 0.25 (Fig.~\ref{fig:lens};
Fig.~\ref{fig:lensprop}; Table~\ref{tab:lens}). For NA 0.5 the spot is diffraction limited (FWHM 1.57 / 1.48~$\upmu$m in
$x$ / $y$, limit 1.59~$\upmu$m) and the DOF is 9.0~$\upmu$m, practically identical to that of the ideal lens
(8.9~$\upmu$m). The model predicts 81.7\% focusing efficiency against 91.8\% for the ideal lens; most of the
difference comes from sampling the phase on the 700~nm grid, not from the atoms' transmission. The FDTD field
reproduces the propagation pattern, the DOF (9.1~$\upmu$m) and the lateral FWHM (1.60 / 1.43~$\upmu$m) within a few
percent, with the focus 0.5~$\upmu$m closer to the lens, and gives 77.3\% focusing efficiency, nearly unchanged between
20, 25 and 30~pixels/$\upmu$m. The 4-point gap is the cost of inter-atom coupling at a steep phase gradient
(4.4 atoms per 2$\pi$ at the lens edge), consistent with the supercell results. For NA 0.25 the agreement is equally
good (DOF 25.7 versus 25.2~$\upmu$m, FWHM 2.53 / 2.50 versus 2.49 / 2.42~$\upmu$m), and FDTD gives a slightly higher
efficiency than the model (90.1\% versus 88.4\%). The long, asymmetric axial profile of this lens and the position of
its intensity maximum, well before the geometric focus (about 29--31 versus 39.3~$\upmu$m), are both expected for a
Fresnel number $R^2/(\lambda f) = 1.7$~\cite{LiWolf1981}; the depth of focus extends from 22 to 48~$\upmu$m.

\begin{table}[t]
\caption{Metalenses assembled from the library ($D = 20.3\um$, $29 \times 29$ atoms): ideal lens, locally periodic
model with angular-spectrum propagation (LPA), and three-dimensional FDTD at 30~pixels/$\upmu$m propagated with the same
code. Focus: maximum of the on-axis intensity, measured from the lens; DOF: full width at half maximum of the on-axis
intensity; FWHM: lateral width at the focus.}
\label{tab:lens}
\centering
\small
\begin{tabular}{llcccccc}
\toprule
NA ($f$) & Model & $T$ & Focus ($\upmu$m) & DOF ($\upmu$m) & FWHM $x$ / $y$ ($\upmu$m) & Focusing eff. & Strehl \\
\midrule
0.50 (17.6~$\upmu$m) & ideal lens & 1     & 16.4 & 8.9  & 1.56 / 1.48 & 91.8\% & 1 \\
                     & LPA        & 0.98  & 16.4 & 9.0  & 1.57 / 1.48 & 81.7\% & 0.88 \\
                     & FDTD       & 0.865 & 15.8 & 9.1  & 1.60 / 1.43 & 77.3\% & -- \\
0.25 (39.3~$\upmu$m) & ideal lens & 1     & 30.9 & 25.6 & 2.52 / 2.48 & 92.6\% & 1 \\
                     & LPA        & 0.98  & 30.7 & 25.7 & 2.53 / 2.50 & 88.4\% & 0.94 \\
                     & FDTD       & 0.963 & 28.9 & 25.2 & 2.49 / 2.42 & 90.1\% & -- \\
\bottomrule
\end{tabular}
\end{table}

\section{Process record, detected issues and cost}
\label{sec:record}
\paragraph{What the agent contributed.} The solver and optimizer are conventional; what changed is the systematic
application of checks that a careful designer would make but rarely makes for every run. Table~\ref{tab:issues} lists
the issues that were detected by the analysis checklist or by the validation stage during this work, together with
the corrective action proposed by the agent. Several of them, including the miscalibrated threshold robustness, the
local optima of a single initial shape and the RCWA truncation error in one polarization, would have led to wrong
conclusions if only the nominal FoM had been inspected. The fixed protocol also made the work reproducible: every
decision is visible as a configuration file, a job list and a report.

\begin{table}[t]
\caption{Issues detected during this work and the corresponding actions.}
\label{tab:issues}
\centering
\small
\begin{tabular}{p{0.30\linewidth}p{0.28\linewidth}p{0.34\linewidth}}
\toprule
Observation & Detected by & Action \\
\midrule
100~nm Si layer: about 50$^\circ$ phase range & feasibility check & disk sweep, $h = 800$~nm agreed with the user \\
Best nominal grating loses 20 points at $\pm$10~nm & exact edge-shift report & geometric robust optimization \\
Threshold robustness acts as $\pm$17--20~nm & grey-level and shift reports & exact geometric erosion/dilation \\
Late activation of robustness stalls & convergence history & robust from iteration 0 \\
Phase levels trapped from one initial disk & local-optima check & sweep-informed multi-shape starts \\
y polarization loses $\sim$30\% to reflection & per-order loss analysis & proposed anti-reflection or bilayer \\
FDTD instabilities at 50--60 px/$\upmu$m & failed validation jobs & Courant 0.35, bilinear material grid \\
RCWA vs FDTD gap of 3.6 points (y pol.) & cross-solver verdict & higher RCWA truncation; verdict \emph{agree} \\
\bottomrule
\end{tabular}
\end{table}

\paragraph{Process record.} Table~\ref{tab:record} summarizes the activity of the agent over the work reported
here, as logged by the execution bridge and by the run directories. All requests passed through the whitelisted
bridge; no credential was handled by the agent. The number of language-model calls and the human time spent reviewing
proposals were not recorded and are therefore not reported; instrumenting the bridge to log them is straightforward
and is planned for the systematic evaluation discussed below.

\begin{table}[t]
\caption{Process record of the agent for the work reported in this paper.}
\label{tab:record}
\centering
\small
\begin{tabular}{p{0.55\linewidth}p{0.35\linewidth}}
\toprule
Quantity & Value \\
\midrule
Bridge requests executed (whitelisted actions) & 84 \\
SLURM jobs submitted & 134 \\
\quad GPU optimization array tasks / single optimizations & 115 / 3 \\
\quad GPU environment checks & 2 \\
\quad FDTD validation jobs (CPU, MPI) & 14 \\
Optimization runs with a complete summary & 96 \\
\quad meta-atoms / metagratings and test examples / supercells & 77 / 16 / 3 \\
Total GPU wall time of the optimization runs & 11.2~h \\
Issues detected and acted upon (Table~\ref{tab:issues}) & 8 \\
Credentials handled by the agent & none \\
\bottomrule
\end{tabular}
\end{table}

\paragraph{Cost.} On one V100S GPU a single metagrating optimization takes 6--19~min (non-robust and robust), a
meta-atom 5~min, a 12-atom supercell refinement 50~min; the 48 atom optimizations of the final library took 4.5~h.
FDTD validation of a periodic design costs minutes to tens of minutes on one 48-core node, and a 20~$\upmu$m lens
1--2~h at 30~pixels/$\upmu$m. The agent reports such estimates before submitting large requests.

\paragraph{Reproducibility.} The repository contains the code, the agent instructions (\texttt{AGENTS.md}), a command
reference (Appendix~\ref{app:commands}) and all runs discussed here as a worked example: a new user can reproduce
the main numbers on a laptop CPU in about ten minutes, without a cluster, and obtain the validation verdicts of
Table~\ref{tab:valid} from the shipped data.

\section{Positioning within AI-driven photonic design frameworks}
\label{sec:positioning}
\paragraph{Positioning within AI-driven photonic design frameworks.} Table~\ref{tab:compare} places the present
pipeline among the three categories introduced in the main text. LLM scripting
assistants~\cite{Kim2025LLM,Lu2025} lower the entry barrier but leave the choice of solver, the verification and
the interpretation of results to the user. Surrogate and generative approaches~\cite{Ma2024,Jiang2021,Lupoiu2025}
make each evaluation cost milliseconds, which enables real-time conversation, massive multi-start batches and large
two-dimensional devices; their accuracy is limited by the training distribution (test errors of a few percent to
about 10\% for MetaChat~\cite{Lupoiu2025}) and they do not by themselves certify a design. A rigorous solver costs
seconds to minutes per evaluation, but every number it returns can be checked for convergence and cross-checked with
an independent method, in three dimensions, for both polarizations and under fabrication perturbations. The
approaches are complementary: a surrogate could propose candidates that the protocol of this work then optimizes
rigorously, robustifies and validates. Conversely, MetaChat evaluates the agent itself on a benchmark and across
several language models, which we have not done: our evidence for the role of the agent is the process record of
Tables~\ref{tab:issues} and \ref{tab:record}, and a quantitative evaluation of protocol adherence and decision
quality across agents is an important direction for future work.

\begin{table}[t]
\caption{Positioning of the present pipeline among AI-driven photonic design approaches, as described in the cited
works. Efficiencies are not compared because devices, dimensionality and definitions differ.}
\label{tab:compare}
\centering
\footnotesize
\begin{tabular}{>{\raggedright\arraybackslash}p{0.15\linewidth}>{\raggedright\arraybackslash}p{0.23\linewidth}>{\raggedright\arraybackslash}p{0.25\linewidth}>{\raggedright\arraybackslash}p{0.25\linewidth}}
\toprule
 & (A) LLM scripting assistants~\cite{Kim2025LLM,Lu2025} & (B) Surrogate/generative frameworks~\cite{Ma2024,Jiang2021,Lupoiu2025} & (C) This work \\
\midrule
Solver in the loop & user-chosen solver, called from generated code & trained network (e.g., FiLM WaveY-Net on FDFD data in MetaChat) & differentiable 3D RCWA (TORCWA), no training \\
Dimensionality, polarization & as set up by the user & 1D thin films (OptoGPT); 2D, TM (MetaChat) & 3D, both polarizations (unpolarized FoM) \\
Speed & that of the solver & milliseconds per evaluation & about 1~s per evaluation; minutes to hours per design \\
Accuracy and verification & left to the user & bounded by the training data; full-wave check of selected designs & convergence check on every run; mandatory 3D FDTD referee with explicit verdict \\
Fabrication robustness & not addressed & minimum feature via neural SDF (MetaChat) & exact $\pm d$~nm edge shifts in the objective; feature-size report \\
Compute and security & local & local GPU server & local GPU/CPU or SLURM cluster via a credential-free whitelisted bridge \\
Agent evaluation & examples & 101-problem benchmark, several LLMs (MetaChat) & process record (Tables~\ref{tab:issues}, \ref{tab:record}) \\
\bottomrule
\end{tabular}
\end{table}

\section{Command reference}
\label{app:commands}
Table~\ref{tab:commands} lists the commands of the pipeline, when to use them and what they produce. All commands
run from the repository root; Python commands run either locally or through the bridge.

\begin{table}[h]
\caption{Commands of the pipeline.}
\label{tab:commands}
\centering
\footnotesize
\begin{tabular}{p{0.36\linewidth}p{0.22\linewidth}p{0.34\linewidth}}
\toprule
Command & When & Result \\
\midrule
\multicolumn{3}{l}{\emph{Execution bridge} (\texttt{hpc/bridge.sh})} \\
\texttt{login}, \texttt{push}, \texttt{setup}, \texttt{setup-meep} & once per site or session & credentials refreshed by the user; code and environments on the target \\
\texttt{check} & new site & GPU visible, eigendecomposition timing \\
\texttt{submit <config> [args]} & one optimization & one run directory \\
\texttt{array <list>} & sweeps, multi-starts, libraries & one task per line (\texttt{run\_topopt.py} arguments or a script) \\
\texttt{meep <list>} & FDTD validation & one MPI run per line, on a CPU node \\
\texttt{watch} & agent mode & executes queued one-line requests (whitelist) \\
\texttt{status}, \texttt{log}, \texttt{cancel}, \texttt{usage} & monitoring & queue, logs, consumed budget \\
\midrule
\multicolumn{3}{l}{\emph{Design}} \\
\texttt{run\_topopt.py --config c.json [--set k=v] [--init f]} & any periodic cell & summary with all orders, convergence, robustness, features \\
\texttt{disk\_sweep.py} & before a library & $T$ and phase versus radius and height \\
\texttt{tools/make\_lut\_sweep.py} & library job list & one run per phase and initial shape \\
\texttt{select\_atoms.py} & after the atom runs & library: best worst-case atom per phase \\
\texttt{supercell.py --library L --levels N} & deflector of given angle & full-supercell RCWA, library prediction, robustness \\
\texttt{metalens.py --library L --diameter-um D --na NA} & metalens & layout, focusing model, efficiency, Strehl \\
\texttt{evaluate.py} & fair comparison & re-evaluation at higher truncation with exact edge shifts \\
\midrule
\multicolumn{3}{l}{\emph{Validation}} \\
\texttt{validation/validate.py plan --runs \dots{} --level quick|full} & validation round & GPU and FDTD job lists, time estimate \\
\texttt{validation/validate.py report --runs \dots} & after the round & verdict per design, shift and polarization; rank agreement \\
\texttt{validation/rcwa\_reference.py}, \texttt{torcwa\_fields.py} & RCWA side & per-order references; near fields \\
\texttt{validation/meep\_validate.py}, \texttt{meep\_lens.py} & FDTD side & per-order efficiencies; lens focus \\
\texttt{validation/compare\_fields.py} & field comparison & near fields, beam and order spectra of both solvers \\
\texttt{paper/make\_figures.py} & after the runs & all figures of this paper from the run directories \\
\bottomrule
\end{tabular}
\end{table}

\clearpage
\section{Transfer to a new problem: TM quasi-BIC at 1550 nm}
\label{sec:qbic}

To test whether the protocol carries over to problems outside the case studies of the main text, a new agent session, started without prior context on the same repository and executing on the
SLURM cluster through the bridge, received the following request "With full freedom, staying on Si/SiO$_2$ around 1550~nm (1540--1560~nm), optimize a structure that shows, in TM polarization, a quasi-BIC mode with a $Q$ of about 1000.'' Refractive indices
were constant ($n_\mathrm{SiO_2} = 1.444$); light is incident from air at normal incidence.

\paragraph{Design route.} The pipeline had no objective for the quality factor, so the agent added a spectral scanning tool based on the same RCWA solver (TORCWA, double precision), which computes the zero-order transmission and reflection and extracts $\lambda_0$ and $Q = \omega_0/2\gamma$ from a complex Fano fit. It chose a dimerized one-dimensional grating, two Si bars per period: in the symmetric cell the mode is a symmetry-protected bound state in the continuum, invisible in the spectra, and a perturbation of amplitude $\alpha$ makes it radiate with $Q \propto \alpha^{-2}$.
Two perturbations were compared (Table~\ref{tab:qbic_pert}); the lateral shift $s$ of one bar was selected because it requires about twice the perturbation of a width difference for the same $Q$, and because a uniform width bias changes neither perturbation, so it shifts $\lambda_0$ but leaves $Q$ almost unchanged.

\begin{table}[h]
\centering
\caption{Quality factor versus perturbation for the dimerized grating ($\Lambda = 900$~nm,
$h = 350$~nm, bar width 220~nm, RCWA order 40). Width difference $d$ and lateral shift $s$ give two
different quasi-BIC modes, both following $Q \propto \alpha^{-2}$.}
\label{tab:qbic_pert}
\begin{tabular}{lccc}
\toprule
Perturbation & Amplitude (nm) & $\lambda_0$ (nm) & $Q$ \\
\midrule
Width difference $d$ & 10 / 20 / 40 & 1433.4 / 1431.6 / 1440.3 & 985 / 252 / 71 \\
Lateral shift $s$    & 20 / 40 / 80 & 1482.1 / 1483.1 / 1488.4 & 1212 / 303 / 76 \\
\bottomrule
\end{tabular}
\end{table}

\begin{figure}[h]
\centering
\includegraphics[width=\linewidth]{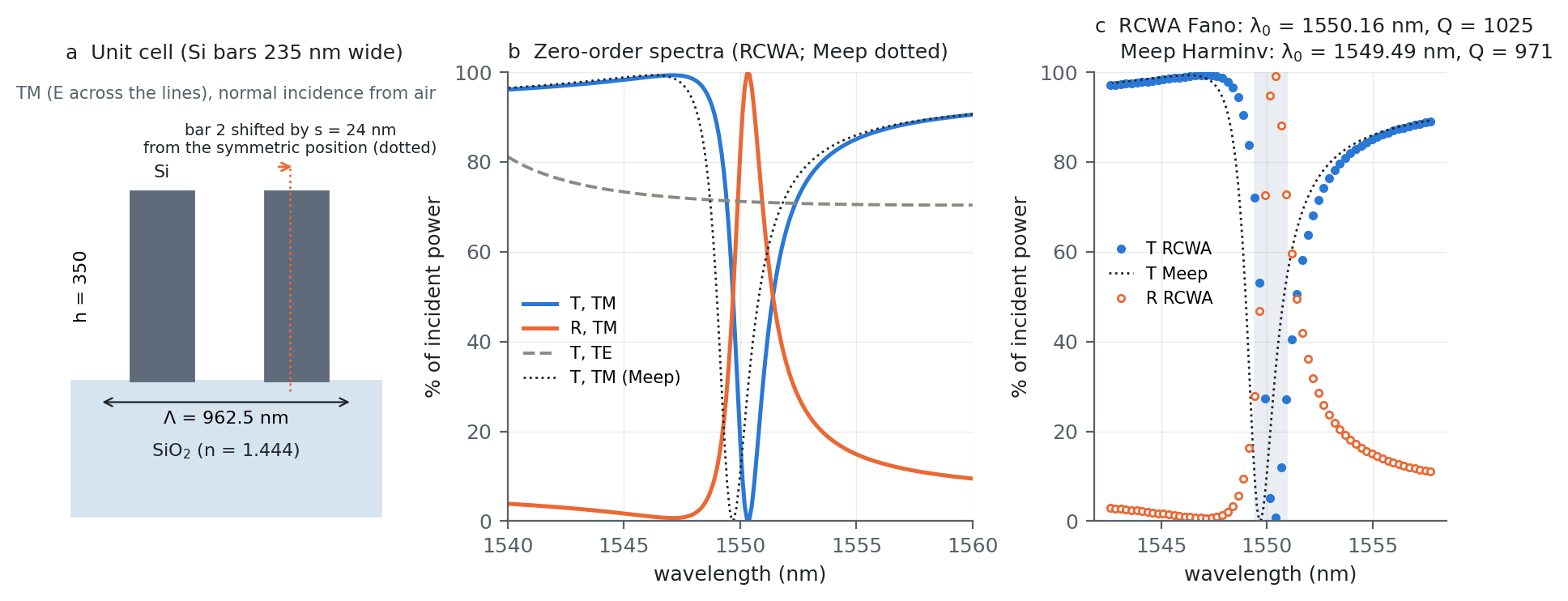}
\caption{TM quasi-BIC at 1550~nm in a dimerized Si grating on SiO$_2$. (a) Unit cell:
$\Lambda = 962.5$~nm, $h = 350$~nm, two 235~nm Si bars, bar 2 shifted by $s = 24$~nm from the
symmetric position. (b) Zero-order transmission and reflection (RCWA, order 90) for TM, transmission
for TE, and transmission from Meep (2D, 320~px/$\upmu$m, dotted). (c) Zoom of the resonance: RCWA
Fano fit $\lambda_0 = 1550.16$~nm, $Q = 1025$; Meep harmonic inversion $\lambda_0 = 1549.49$~nm,
$Q = 971$.}
\label{fig:qbic}
\end{figure}

The resonance was centred by scaling the period and the bar width, and $s$ was set to 24~nm for $Q \approx 1000$. Raising the Fourier truncation from 40 to 60 and 90 orders shifted $\lambda_0$ from
1550.16 to 1548.29 and 1547.35~nm and changed $Q$ from 1002 to 1018 and 1021; the final design was therefore centred at order 90. Final design: $\Lambda = 962.5$~nm, $h = 350$~nm, two 235~nm bars,
bar 2 displaced by $s = 24$~nm, with $\lambda_0 = 1550.16$~nm, $Q = 1025$ and a linewidth of 1.51~nm (Fano fit rms 0.002; $|1 - T - R| < 3\times10^{-10}$). In TM polarization the transmission falls from about 100\% to 0 and the reflection rises to about 100\% at resonance; TE shows no resonance (Fig.~\ref{fig:qbic}). Table~\ref{tab:qbic_fab} gives the sensitivity to fabrication errors: a uniform bias of $\pm10$~nm keeps $\lambda_0$ within 1540--1560~nm, and the thickness is the most critical parameter (1.35~nm of resonance shift per nm).

\begin{table}[h]
\centering
\caption{Sensitivity of the quasi-BIC to fabrication errors (RCWA order 40).}
\label{tab:qbic_fab}
\begin{tabular}{lcc}
\toprule
Error & $\lambda_0$ (nm) & $Q$ \\
\midrule
Nominal                      & 1550.2          & 1002 \\
Uniform width bias $-10$ / $+10$~nm & 1541.0 / 1558.7 & 1291 / 809 \\
Uniform width bias $-5$ / $+5$~nm   & 1545.5 / 1554.9 & 1134 / 888 \\
Width mismatch $\pm5$~nm     & 1550.1 / 1549.2 & 991 / 1016 \\
Shift error $-5$ / $+5$~nm   & 1549.6 / 1549.9 & 1582 / 703 \\
Thickness $-10$ / $+10$~nm   & 1536.8 / 1563.6 & 1035 / 974 \\
\bottomrule
\end{tabular}
\end{table}

\paragraph{Independent FDTD verification.} The agent wrote a two-dimensional Meep check (periodic in $x$, absorbing layers along the propagation axis, source in air) that extracts $\lambda_0$ and $Q$ directly from the time decay of the field with harmonic inversion, without a fit, and computes the transmission spectrum normalized to an empty-cell reference. The first resolution study gave scattered values of $Q$ (Table~\ref{tab:qbic_meep}). The agent traced the scatter to the grid rather than to the physics: because $Q \propto s^{-2}$, any fraction of a pixel of spurious asymmetry matters, and at 80 and 120~px/$\upmu$m the half-period or the period does not span an integer number of grid cells, so the discretized ``symmetric'' cell is no longer symmetric. Only grids commensurate with the cell give consistent values: at 320~px/$\upmu$m Meep gives $\lambda_0 = 1549.49$~nm and $Q = 971$, i.e., 0.7~nm (less than half a linewidth) and 5\% from RCWA, with $\lambda_0$ approaching the RCWA value as the resolution increases. Off resonance the transmission also agrees (96.5\% and 90.6\% at 1540 and 1560~nm, against 96\% and 91\% from RCWA). The two FDTD jobs took about 33~min on one CPU node (about 25 core-hours).

\begin{table}[h]
\centering
\caption{FDTD resolution study of the quasi-BIC (Meep, harmonic inversion, TM). Only resolutions for
which both the period and the half-period span an integer number of grid cells give consistent
results.}
\label{tab:qbic_meep}
\begin{tabular}{cccl}
\toprule
Resolution (px/$\upmu$m) & $\lambda_0$ (nm) & $Q$ & Grid \\
\midrule
80  & 1556.2  & 1292 & half-period off-grid \\
120 & 1555.3  & 1908 & period rounded by the code \\
160 & 1552.71 & 998  & commensurate \\
320 & 1549.49 & 971  & commensurate \\
\bottomrule
\end{tabular}
\end{table}

\paragraph{A discarded variant: reflection quasi-BIC in Si at 532~nm.} 
In the same session the request was extended to a reflection quasi-BIC at 532~nm. Before any optimization the agent noted that crystalline Si absorbs at this wavelength (it assumed $n = 4.15 + 0.044i$, quoted from the literature rather than taken from a database), which limits the attainable $Q$ to a few tens whatever the design. A lossy dimer scan confirmed this: as the shift $s$ grows from 10 to 40~nm the reflection peak rises from 52\% to 80\% while $Q$ falls from about 49 to 23 and the maximum absorption from 58\% to 21\%. Freeform optimizations at 532~nm initially reported narrow reflection peaks (reflectance 64\% - 70\%), but the convergence check showed changes of 44\% - 53\% when the Fourier order was raised from 8 to 12, so these peaks were largely truncation artifacts, consistent with the high index contrast and the short period (300~nm). Converged spectra (orders 16 and 20, differing by at most 3\% in reflectance) showed only a broad, mainly absorbing resonance (absorption 66\% at 526~nm, reflectance at most 45\% at 534~nm), and the Si route was abandoned.

\clearpage
\bibliographystyle{unsrtnat}
\bibliography{refs}